\documentclass[preprint,authoryear,sort&compress,12pt]{elsarticle}  
\usepackage{fullpage}
\usepackage{graphicx}
\usepackage{amssymb}
\usepackage{amsmath}
\usepackage{lineno}
\usepackage{subfig}

\usepackage{color}
\usepackage{soul}
\usepackage{array}
\usepackage{hyperref}
\usepackage[colorinlistoftodos]{todonotes}

\newcommand{\xmb}[1]{\ensuremath{\mathbf{#1}}}
\newcommand{\xmbs}[1]{\ensuremath{\boldsymbol{#1}}}

\newcommand*\patchAmsMathEnvironmentForLineno[1]{%
  \expandafter\let\csname old#1\expandafter\endcsname\csname #1\endcsname
  \expandafter\let\csname oldend#1\expandafter\endcsname\csname end#1\endcsname
  \renewenvironment{#1}%
     {\linenomath\csname old#1\endcsname}%
     {\csname oldend#1\endcsname\endlinenomath}}%
\newcommand*\patchBothAmsMathEnvironmentsForLineno[1]{%
  \patchAmsMathEnvironmentForLineno{#1}%
  \patchAmsMathEnvironmentForLineno{#1*}}%
\AtBeginDocument{%
\patchBothAmsMathEnvironmentsForLineno{equation}%
\patchBothAmsMathEnvironmentsForLineno{align}%
\patchBothAmsMathEnvironmentsForLineno{flalign}%
\patchBothAmsMathEnvironmentsForLineno{alignat}%
\patchBothAmsMathEnvironmentsForLineno{gather}%
\patchBothAmsMathEnvironmentsForLineno{multline}%
}

\journal{}

\begin{document}

\begin{frontmatter}

\title{CFD--DEM Simulations of Current-Induced Dune Formation and Morphological Evolution}

\author{Rui Sun} \ead{sunrui@vt.edu}
\author{Heng Xiao\corref{corxh}} \ead{hengxiao@vt.edu}
\cortext[corxh]{Corresponding author. Tel: +1 540 231 0926}

\address{Department of Aerospace and Ocean Engineering, Virginia Tech, Blacksburg, VA 24060, United
 States}


\begin{abstract}
  Understanding the fundamental mechanisms of sediment transport, particularly those during the
  formation and evolution of bedforms, is of critical scientific importance and has engineering
  relevance. Traditional approaches of sediment transport simulations heavily rely on empirical
  models, which are not able to capture the physics-rich, regime-dependent behaviors of the process.
  With the increase of available computational resources in the past decade, CFD--DEM (computational
  fluid dynamics--discrete element method) has emerged as a viable high-fidelity method for the
  study of sediment transport. However, a comprehensive, quantitative study of the generation and
  migration of different sediment bed patterns using CFD--DEM is still lacking. In this work,
  current-induced sediment transport problems in a wide range of regimes are simulated, including
  `flat bed in motion', `small dune', `vortex dune' and suspended transport.  Simulations are
  performed by using \textit{SediFoam}, an open-source, massively parallel CFD--DEM solver developed
  by the authors. This is a general-purpose solver for particle-laden flows tailed for particle
  transport problems.  Validation tests are performed to demonstrate the capability of CFD--DEM in
  the full range of sediment transport regimes.  Comparison of simulation results with experimental
  and numerical benchmark data demonstrates the merits of CFD--DEM approach. In addition, the
  improvements of the present simulations over existing studies using CFD--DEM are presented. The
  present solver gives more accurate prediction of sediment transport rate by properly accounting
  for the influence of particle volume fraction on the fluid flow. In summary, this work
  demonstrates that CFD--DEM is a promising particle-resolving approach for probing the physics of
  current-induced sediment transport.
\end{abstract}


\end{frontmatter}


\section{Introduction}
\label{sec:intro}

The perpetual motion of water carves the surface of the earth by entraining and carrying sediment
from one location to another, leading to changes of morphology in the ocean and particularly along
the coastline.  Scientists rely on fundamental understanding of sediment transport to explain and
predict the dynamic evolution of the seabed and coastal bathymetry at various spatial and temporal
scales; engineers utilize the understanding of the sediment transport mechanisms to design
better civil defense infrastructure, which mitigates the impact of coastal hazards such as storm
surges and tsunamis on the coastal communities. However, the understanding and prediction of
sediment transport are hindered by the complex dynamics and numerous regimes. Traditional hydro- and
morphodynamic models~\citep{MG-num,lesser00ol,hydraulics99df,warren1992mike} for sediment transport
simulations heavily relied on phenomenological models and empirical correlations to describe
sediment erosion and deposition fluxes~\citep{meyer48fb,rijn84se1}, which lack universal
applicability across different regimes and can lead to large discrepancies in predictions.

With the rapid growth of available computational resources in the past decades, many high-fidelity
models have been proposed, including two-fluid models~\citep{hsu04tp,yu2012pv}, particle-resolving
models~\citep{drake01dp,calantoni04ms,schmeeckle14ns,jiang95tm}, and interface-resolving
models~\citep{kempe12cm,kempe14ot,kidanemariam14id,kidanemariam14dn}. Two-fluid models describe the
particle phase as a continuum and thus need constitutive relations to account for the
particle--particle collisions and fluid--particle interactions. Particle-resolving models explicitly
track the movements of all particles and their collisions, which are thus much more expensive than
two-fluid models. Empirical models are still used to compute the fluid--particle interaction forces.
In interface-resolving models, not only individual particles but also the detailed flows fields
around particle surfaces are fully resolved. Consequently, they are more expensive than
particle-resolving models but require even less empirical modeling.

Particle-resolving models can accurately predict particle phase dynamics such as vertical and
horizontal sorting due to densities, sizes, shapes, which are important phenomena in nearshore
sediment transport. Possibly constrained by computational resources at the time, early
particle-resolving models used highly simplified assumptions for the fluid phase by modeling the fluid
as two-dimensional layers~\citep{drake01dp,jiang95tm}. The number of particles was also limited to a
few thousand particles, and thus the computational domain covers only a few centimeters or less for
particle diameters typical for coastal sediments. As a result, these methods were limited to
featureless bed under specific flow conditions (e.g., intense sheet flow conditions, where the layer
fluid assumption is valid).

\subsection{Simulation of Sediment Transport with Modern CFD--DEM Methodology}

In the past few years, researchers started to use modern, general-purpose particle-resolving solvers
based on Computational Fluid Dynamics--Discrete Element Method (CFD--DEM) to study sediment
transport.  In CFD--DEM, Reynolds Averaged Navier--Stokes (RANS) equations or Large Eddy Simulations
(LES) are used to model the fluid flows, which are coupled with the discrete element method for the
particles. The CFD--DEM has been used extensively in the past two decades in the chemical and
pharmaceutical industry on a wide range of applications such as fluidized beds, cyclone separator,
and pneumatic conveying~\citep{han03DEM,ebrahimi14cfd}. On the other hand, special-purpose codes
have been used to study specific regimes of sediment transport, where solvers are developed based on
and valid for only the sediment transport regime to be studied, e.g., bedload transport under
two-dimensional, laminar flow conditions~\citep{duran12ns}.  However, the use of modern,
general-purpose CFD--DEM solvers as those used in chemical engineering applications to simulate
sediment transport is only a recent development in the past few years. In his pioneering work,
\cite{schmeeckle14ns} used an open-source CFD--DEM solver~\citep{goniva09tf,kloss12ma} to study
suspended sediment transport. {\color{black} The merits and significance of Schmeecle's pioneering
work are summarized as follows: (1) It is the
first work done by using modern CFD--DEM solver in the simulation of sediment transport, especially
in the suspended sediment transport regime; (2) Rich data sets are obtained by the CFD--DEM solver
that are very difficult to obtain in the field or the laboratory; (3) Several questions of the
mechanics of sediment transport are answered, including the mechanisms of saltation and entrainment;
(4) Interesting and insightful phenomena are observed, including the increase of bed friction at the
transition of suspension. However, a theoretical limitation of his work is that the influence of
particle volume fraction on the fluid flow is not considered, since the volume fraction does not
appear in the fluid continuity equation (see Eq.~(1) in \cite{schmeeckle14ns}).} This choice was
likely made to avoid the destabilizing effects of the volume fraction on the LES equations.
Moreover, the fluid--particle drag law adopted in his work does not explicitly account for the
volume fraction.  Consequently, the drag law he used is not able to represent the varying shielding
effects of particles under different particle loading conditions.  This effect is important in
particle-laden flows where the flow field has disparate distributions of particle loadings from very
dilute to very dense, which is the consensus of the CFD--DEM
community~\citep{tsuji93,kafui02,fengyu07} in simulating industrial particle-laden flows. Finally,
the study by \citet{schmeeckle14ns} focused on suspended sediment on featureless beds with
comparison of sediment transport rates to empirical formulas in the literature.  Many other regimes
of sediment transport such as bedload transport as well as more complex patterns such as the
formation and evolution of bedforms are still yet to be studied.  \cite{arolla15tm} studied
the transport of cuttings particles in a pipe with CFD--DEM, where a volume-filtered LES approach is
used to model the fluid flow~\citep{capecelatro13ae}. The emergence of small dunes and sinusoidal
dunes from an initially flat particle bed under different flow velocity are observed, demonstrating
the capability of CFD--DEM in predicting the stability characteristics of sediment beds.  However,
quantitative comparisons with experimental data are limited to a few integral quantities such as
holding rate, and a more detailed validation with experimental or numerical benchmark data were not
performed. In summary, while a few researchers have made attempts in using CFD--DEM to study
sediment transport and have obtained qualitatively reasonable predictions, a rigorous, comprehensive
study of sediment transport in a wide range of regimes with detailed quantitative comparisons with
benchmark data is still lacking.  This study aims to bridge this gap by tackling the unique
challenges for the CFD--DEM posed by the physical characteristics of sediment transport problems,
which are detailed below.


\subsection{Unique Challenges of Sediment Transport with CFD--DEM}

Given the decades of experiences of using CFD--DEM in chemical engineering applications, one may
expect that all these experiences should be straightforwardly transferable to simulations of
sediment transport. Unfortunately, this is not the case.  First, most of the critical phenomena such
as incipient motion, entrainment, suspension, and mixing of suspended sediments with water occur in
a boundary layer near the interface of the fluid and the sediment bed.  Adequately resolving the
flow features within the boundary layer such as the mean velocity gradient, shear stress, and
turbulent coherent structures is essential for capturing the overall dynamics of fluid and particle
flows. In contrast, in fluidized bed applications, the dynamics of the fluids and particles in the
entire bed are of equal importance.  Accurately resolving the boundary layer features poses both
theoretical and practical challenges for CFD--DEM. This is because the characteristic length scales
of the flow can be comparable to or smaller than the particle diameters, but the CFD--DEM describes
the fluid flows with \emph{locally averaged} Navier--Stokes equations, which are only valid at
scales much larger than the particle size~\citep{anderson67}.  Moreover, since the carrier phase
(water) and the dispersed phase (particles) have comparable densities in sediment transport, many
effects that are negligible in gas--solid flows such as added mass effects and lubrication are
important sediment transport. In comparison, the density of the carrier phase (air or other gases)
in gas-solid flows is two orders of magnitude smaller than that of the particles. Consequently, the
fluid--particle interactions are dominated by the drag forces, while the other forces mentioned
above are of secondary importance and can be neglected~\citep{zhou11dp}.

In this work, we demonstrate that CFD--DEM is able to capture the essential features of sediment
transport in various regimes with a small fraction of the computational cost of interface-resolved
models. On the other hand, detailed features in the bed dynamics in the turbulent flows are
reproduced correctly, which is beyond the reach of lower fidelity models such as two-fluid models or
phenomenological model based morphodynamic simulations. Furthermore, we demonstrate that improved
results can be obtained by properly accounting for the effects of particle volume fraction on the
fluid dynamics and the fluid-particle interaction forces.  Therefore, when properly used,
CFD--DEM can be a powerful and practical tool to probe the fundamental dynamics of sediment
transport across a wide range of regimes.

The rest of the paper is organized as follows. Section~2 presents the theoretical framework of
CFD--DEM approach. The technique adopted to address the difficulty of comparable scales between the
boundary layer and the particle sizes in sediment transport is introduced. Section~3 summarizes the
implementation of the CFD--DEM solver SediFoam and the numerical methods used in the
simulations. The results are presented and discussed in Section~4~and~5, respectively. Finally,
Section~6 concludes the paper.

\section{Methodology}
\label{sec:cfddem}

\subsection{Mathematical Model of Particle Motion}
\label{sec:dem-particle}

In CFD--DEM, the translational and rotational motion of each particle is calculated
based on Newton's second law as the following equations~\citep{cundall79,ball97si}: 
\begin{subequations}
 \label{eq:newton}
 \begin{align}
  m \frac{d\xmb{u}}{dt} &
  = \xmb{f}^{col} + \xmb{f}^{fp} + m \xmb{g} \label{eq:newton-v}, \\
  I \frac{d\xmbs{\Psi}}{dt} &
  = \xmb{T}^{col} + \xmb{T}^{fp} \label{eq:newton-w},
 \end{align}
\end{subequations}
where \( \xmb{u} \) is the velocity of the particle; $t$ is time; $m$ is particle mass;
\(\xmb{f}^{col} \) represents the contact forces due to particle--particle or particle--wall
collisions; \(\xmb{f}^{fp}\) denotes fluid--particle interaction forces; \(\xmb{g}\) denotes body
force. Similarly, \(I\) and \(\xmbs{\Psi}\) are angular moment of inertia and angular velocity of
the particle; \(\xmb{T}^{col}\) and \(\xmb{T}^{fp}\) are the torques due to contact forces and
fluid--particle interactions, respectively.  To compute the collision forces and torques, the
particles are modeled as soft spheres with inter-particle contact represented by an elastic spring
and a viscous dashpot.

\subsection{Locally-Averaged Navier--Stokes Equations for Fluids}
\label{sec:lans}

The fluid phase is described by the locally-averaged incompressible Navier--Stokes equations.
Assuming constant fluid density \(\rho_f\), the governing equations for the fluid
are~\citep{anderson67,kafui02}:
\begin{subequations}
 \label{eq:NS}
 \begin{align}
  \nabla \cdot \left(\varepsilon_s \xmb{U}_s + {\varepsilon_f \xmb{U}_f}\right) &
  = 0 , \label{eq:NS-cont} \\
  \frac{\partial \left(\varepsilon_f \xmb{U}_f \right)}{\partial t} + \nabla \cdot \left(\varepsilon_f \xmb{U}_f \xmb{U}_f\right) &
  = \frac{1}{\rho_f} \left( - \nabla p + \nabla \cdot \xmbs{\mathcal{R}} + \varepsilon_f \rho_f \xmb{g} + \xmb{F}^{fp}\right), \label{eq:NS-mom}
 \end{align}
\end{subequations}
where \(\varepsilon_s\) is the solid volume fraction; \( \varepsilon_f = 1 - \varepsilon_s \) is the
fluid volume fraction; \( \xmb{U}_f \) is the fluid velocity. The terms on the right hand side of
the momentum equation are: pressure gradient \(\nabla p\), divergence of the stress tensor \(
\xmbs{\mathcal{R}} \) (including viscous and Reynolds stresses), gravity, and fluid--particle
interactions forces, respectively. {\color{black} In the present study, we used large-eddy simulation
to resolve the flow turbulence in the computational domain.  We applied the one-equation eddy
viscosity model proposed by~\cite{yoshizawa85sd} as the sub-grid scale~(SGS) model.} The Eulerian
fields $\varepsilon_s$, $\xmb{U}_s$, and $\xmb{F}^{fp}$ in Eq.~(\ref{eq:NS}) are obtained by
averaging the information of Lagrangian particles.

\subsection{Fluid--Particle Interactions}
\label{sec:fpi}
The fluid-particle interaction force \(\xmb{F}^{fp}\) consists of buoyancy \( \xmb{F}^{buoy} \),
drag \( \xmb{F}^{drag} \), lift force \(\xmb{F}^{lift}\), and added mass force \(\xmb{F}^{add}\).
Although the lift force and the added mass force are usually ignored in CFD--DEM simulations, they
are important in the simulation of sediment transport. 

The drag on an individual particle $i$ is formulated as:
\begin{equation}
  \mathbf{f}^{drag}_i = \frac{V_{p,i}}{\varepsilon_{f, i} \varepsilon_{s, i}} \beta_i \left( \mathbf{u}_{p,i} -
  \mathbf{U}_{f, i} \right),
  \label{eqn:particleDrag}
\end{equation}
where \( V_{p, i} \) and \( \mathbf{u}_{p, i} \) are the volume and the velocity of particle $i$,
respectively; \( \mathbf{U}_{f, i} \) is the fluid velocity interpolated to the center of particle
$i$; \( \beta_{i} \) is the drag correlation coefficient which accounts for the presence of other
particles. The drag force model proposed by~\cite{mfix93} is applied to the present simulations. The
lift force on a spherical particle is modeled as~\citep{saffman65th,rijn84se1}:
\begin{equation}
  \mathbf{f}_{i}^{lift} = C_{l} \rho_f \nu^{0.5} d_p^{2} \left( \mathbf{u}_{p,i} - \mathbf{U}_{f,i}
  \right) \boldsymbol{\times} \nabla \mathbf{U}_{f,i},
  \label{eqn:particleLift}
  \end{equation}
where $\boldsymbol{\times}$ indicates the cross product of two vectors; $d_p$ is the diameter of the
particle; $C_{l} = 1.6$ is the lift coefficient. The added mass force is modeled as:
\begin{equation}
  \mathbf{f}_{i}^{add} = C_{add} \rho_f V_{p,i} \left( \frac{\mathrm{D}\mathbf{u}_{p,i}}{\mathrm{D}t}
  - \frac{\mathrm{D}\mathbf{U}_{f,i}}{\mathrm{D}t} \right),
  \label{eqn:particleAddedMass}
  \end{equation}
where $C_{add} = 0.5$ is the coefficient of added mass.

\section{Implementations and Numerical Methods}
\label{sec:num-method}
The hybrid CFD--DEM solver \textit{SediFoam} is developed based on two state-of-the-art open-source
codes in their respective fields, i.e., a CFD platform OpenFOAM (Open Field Operation and
Manipulation) developed by \citet{openfoam} and a molecular dynamics simulator LAMMPS (Large-scale
Atomic/Molecular Massively Parallel Simulator) developed at the Sandia National
Laboratories~\citep{lammps}. The LAMMPS--OpenFOAM interface is implemented for the communication of
the two solvers. The solution algorithm of the fluid solver in \textit{SediFoam} is partly based on
the work of~\cite{rusche03co} on bubbly two-phase flows. The code is publicly available at
https://github.com/xiaoh/sediFoam under GPL license.  Detailed introduction of the implementations are
discussed in~\cite{sun2016sedi}.

The fluid equations in~(\ref{eq:NS}) are solved in OpenFOAM with the finite volume method
\citep{jasak96ea}. The discretization is based on a collocated grid, i.e., pressure and all velocity
components are stored in cell centers. PISO (Pressure Implicit Splitting Operation) algorithm is
used to prevent velocity--pressure decoupling~\citep{issa86so}. A second-order central scheme is
used for the spatial discretization of convection terms and diffusion terms. Time integrations are
performed with a second-order implicit scheme. An averaging algorithm based on diffusion is
implemented to obtain smooth $\varepsilon_s$, $\xmb{U}_s$ and $\xmb{F}^{fp}$ fields from discrete
sediment particles~\citep{sun14db1, sun14db2}. In the averaging procedure, the diffusion equations
are solved on the CFD mesh. A second-order central scheme is used for the spatial discretization of
the diffusion equation; a second-order implicit scheme is used for the temporal integration.

\section{Results}
\label{sec:simulations}
Simulations are performed using CFD--DEM for three representative sediment transport
problems: `flat bed in motion', generation of dunes, and suspended sediment transport.  The
objective of the simulations is to show the capability of CFD--DEM for different sediment
transport regimes. The first two simulations aim to demonstrate that CFD--DEM can capture the
features of sediment patterns with a small fraction of the computational cost of interface-resolved
method. Therefore, the results obtained are validated with both the numerical benchmark data and
experimental results. The purpose of the third simulation is to show the capability of CFD--DEM in
`suspended load' regime at high Reynolds number. The results obtained in `suspended
load' regime are validated using experimental data.

The numerical setup of the simulations is detailed in Section~\ref{sec:run-setup}.  The study of
sediment transport in `flat bed in motion' regime is presented in Section~\ref{sec:run1-flat}. The
generation of `small dune' and `vortex dune' is discussed in Section~\ref{sec:run2-dune}.
Section~\ref{sec:run3-suspend} details the study of sediment transport in `suspended particle'
regime.

\subsection{Numerical Setup}
\label{sec:run-setup}
\begin{figure}[htbp]
  \centering
  \subfloat[]{
    \label{fig:layour-case1}
    \includegraphics[width=0.45\textwidth]{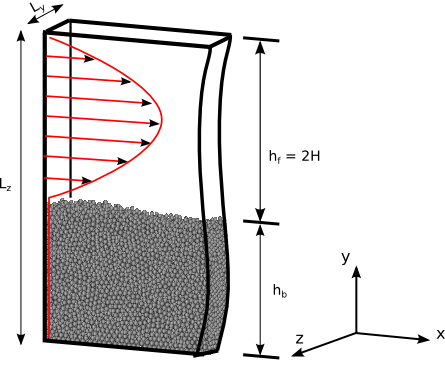}
  }
  \hspace{0.001\textwidth}
  \subfloat[]{
    \label{fig:layout-case3}
    \includegraphics[width=0.45\textwidth]{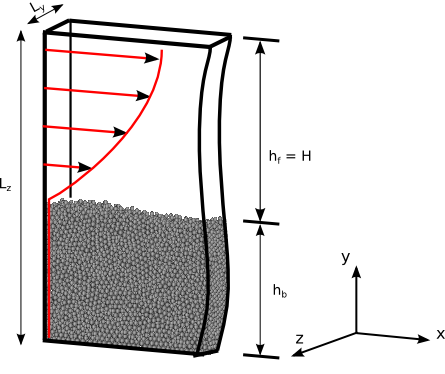}
  }
  \caption{The numerical setup in the simulations. Panel (a) demonstrates the boundary conditions
  used in case 1 and 2a; Panel (b) demonstrates the boundary conditions used in case 2b and 3.}
  \label{fig:layout-st2}
\end{figure}

The numerical tests are performed using a periodic channel. The shape of the computational domain
and the coordinates system are shown in Fig.~\ref{fig:layout-st2}. The Cartesian coordinates $x$,
$y$, and $z$ are aligned with the streamwise, vertical, and lateral directions.  The parameters used
are detailed in Table~\ref{tab:param-sedi}. The numbers of sediment particles range from 9,341 to
330,000 for sediment transport problems of different complexities.

\begin{figure}[htbp]
  \centering
  \includegraphics[width=0.8\textwidth]{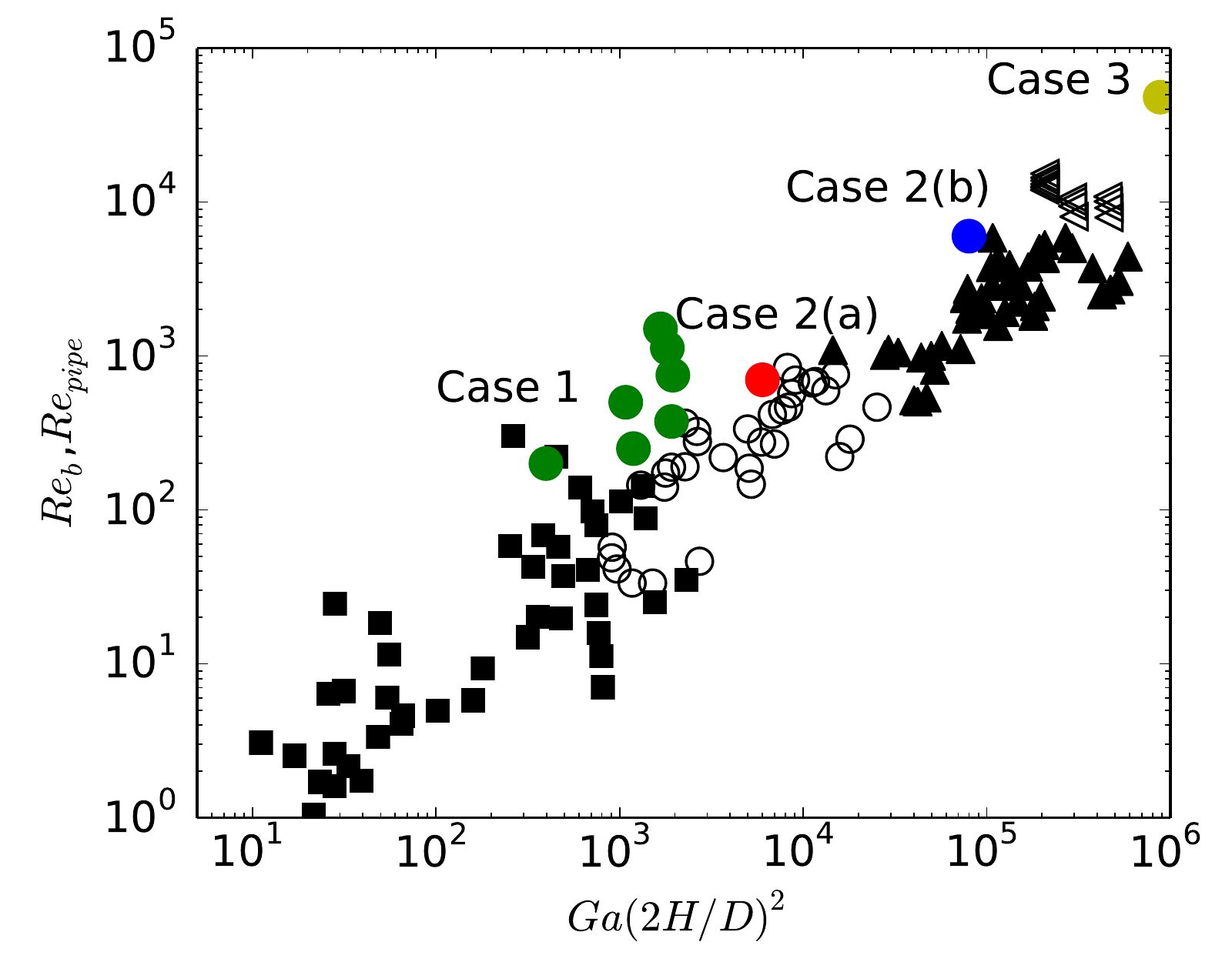}
  \caption{Different patterns observed in the experiments: `flat bed in motion' ($\blacksquare$);
  `small dunes' ($\bigcirc$); `vortex dunes' ($\blacktriangle$); `sinusoidal dunes' ($\lhd$). The
  green diamonds refer to Case 1; the red triangle refers to Case 2a; the blue triangle refers to Case
  2b; the yellow circle refers to Case 3.}
  \label{fig:dune-regime}
\end{figure}

CFD--DEM is used to study the evolution of different dunes according to the regime map in
Fig.~\ref{fig:dune-regime}. This is to demonstrate the capability of CFD--DEM in the prediction of
dune migration. It can be seen from the regime map that the dune height increases with Galileo
number, which is due to the increase of particle inertia. Simulations at different Galileo numbers
are performed to show that CFD--DEM is able to predict the generation of both `small dune' and
`vortex dune'. {\color{black} It can be also seen in Fig.~\ref{fig:dune-regime} that the size of the
dunes is growing from `small dune' to `vortex dune' then to `sinusoidal dune' with the increase of
Reynolds number. However, the influence of Reynolds number to the dune generation is smaller than
that of Galileo number.}

The geometry of different numerical tests are shown in Fig.~\ref{fig:layout-st2}. {\color{black} The
boundary conditions in both $x$- and $z$-directions are periodic in all cases. For the pressure
field, zero-gradient boundary condition is applied in $y$-direction. However, there are slight
differences in the boundary condition for the velocity field. In Case 1 and 2a, the flow is bounded
in the vertical direction by two solid walls and no-slip boundary condition is applied.  On the
other hand, in Case 2b and 3, the simulations are performed in open channels.  In the open channel,
no-slip wall is applied at the bottom while free-slip condition is applied on the top. } The CFD
mesh is refined at the near-wall region and the particle-fluid interface in the vertical ($y$-)
direction to resolve the flow at the boundary layer. {\color{black} Since the CFD mesh is refined and
smaller than the size of sediment particle, a diffusion-based averaging algorithm proposed by the
authors~\citep{sun14db1, sun14db2} is applied to average the quantities (volume fraction, particle
velocity, fluid-particle interaction force) of Lagrangian particles to Eulerian mesh. }The bandwidth
$b$ used in the averaging procedure is $4d_p$ in $x$- and $z$- directions and $2d_p$
in $y$-direction. To model the no-slip boundary condition of sediment particles, an artificial rough
bottom is applied using three layers of fixed sediment particles. The fluid flow is driven by a
pressure gradient to maintain a constant flow rate $q_f$. {\color{black} To resolve the collision
between the sediment particles, the contact force between sediment particles is computed with a
linear spring-dashpot model. In this model, the normal elastic contact force between two particles
is linearly proportional to the overlapping distance~\citep{cundall79}. The stiffness, the
restitution coefficient, and the friction coefficient are detailed in Table~\ref{tab:param-sedi}.
The time step to resolve the particle collision is 1/50 the contact time to avoid particle
inter-penetration~\citep{sun07ht}.}

The initialization of the numerical tests follows the numerical benchmark~\citep{kidanemariam14dn}
using direct numerical simulations (DNS).  The initial positions of the particles are determined in
a separated simulation of particle settling without considering the hydrodynamic forces. In the
particle settling simulation, particles fall from random positions under gravity with inter-particle
collisions. To initialize the turbulent flow in Case 2b and 3, the simulations first run 20
flow-through times with all particles fixed at the bottom.

\begin{table}[!htbp]
  \caption{Parameters used in different simulations of sediment transport.}
 \begin{center}
 \begin{tabular}{lcccc}
   & Case 1 & Case 2a & Case 2b & Case 3\\
   \hline
   bed dimensions & & & \\
   \hline
   \qquad width $(L_x)$~(mm)        & 16    & 156   & 156   & 120   \\
   \qquad height $(L_y)$~(mm)       & 12--16    & 38.4  & 16.7  & 40    \\
   \qquad transverse thickness  $(L_z)$~(mm)& 8     & 8     & 40    & 60    \\
   mesh resolutions                         &       &       &       &\\
   \qquad width $(N_x)$                     & 16    & 90    & 120   & 120   \\
   \qquad height $(N_y)$                    & 100   & 100   & 80    & 65    \\
   \qquad transverse thickness $(N_z)$      & 4     & 6     & 40    & 60    \\
   particle properties  & & &\\
   \qquad total number          & 9341 & 79,000 & 263,000 & 330,000 \\
   \qquad diameter $d_p$~(mm)   & \multicolumn{4}{ c }{0.5} \\
   \qquad density $\rho_s$~($\mathrm{kg/m^3})$  & $2.5\times10^3$ 
                    & $2.5\times10^3$ & $2.5\times10^3$ & $2.65\times10^3$  \\
   \qquad particle stiffness coefficient~(N/m)  & 20 & 200 & 200 & 200 \\
   \qquad normal restitution coefficient & 0.3 & 0.3 & 0.3 & 0.01 \\
   \qquad coefficient of friction & 0.4 & 0.4 & 0.4 & 0.6 \\
   fluid properties & \\
   \qquad density $\rho_f$~($\mathrm{kg/m^3})$ & \multicolumn{4}{ c }{$1.0\times10^3$} \\
   \qquad viscosity~($\mathrm{m^2/s}$)  & $5.0 \times 10^{-6}$ & $1.8 \times 10^{-5}$ 
                                        & $1.5 \times 10^{-6}$ & $1.0 \times 10^{-6}$ \\ 
   \qquad mean velocity~(m/s)   & 0.12--0.67   & 0.46 & 0.34 & 0.8--1.2 \\
   non-dimensional numbers& \\
   \qquad bulk Reynolds number $Re_b$ & 180--1500 & 840 & 6500 & 48000 \\
   \qquad Galileo number $Ga$ & 8.6 & 2.4 & 28.4 & 42.9\\
   \hline
  \end{tabular}
 \end{center}
 \label{tab:param-sedi}
\end{table}

There are several dimensionless numbers to describe the subaqueous sediment transport.  The Galileo
number, or the particle Reynolds number, is defined as $Ga = u_g d_p/\nu$, where $u_g =
((\rho_p/\rho_f - 1)|\xmb{g}|d_p)^{1/2}$. The bulk Reynolds number $Re_{bulk}$ is given by:
\begin{equation}
  Re_{bulk} = \frac{q_f}{\nu} = \frac{2H u_b}{\nu},
  \label{eqn:Re_bulk}
\end{equation}
where $q_f$ is the fluid flow rate (note that the volumetric flow rate is divided by the area of the
horizontal plane); $u_b$ is the bulk velocity; $H$ is the equivalent boundary layer thickness. As
shown in Fig.~\ref{fig:layout-st2}, the boundary layer thickness $H = h_f/2$ in Case 1 and 2a, while
$H = h_f$ in Case 2b and 3. Because the thickness of the boundary layer can be influenced by the
height of the sediment bed, there are slight differences in the Reynolds number obtained from the
present simulations and the numerical benchmark~\citep{kidanemariam14id, kidanemariam14dn}. The fluid
height $h_f$ is defined as $h_f = L_y - h_b$, where $L_y$ is the height of the computational domain;
$h_b$ is the height of sediment bed that is the spatially averaged vertical location of
$\langle\varepsilon_s\rangle = 0.10$.  The Shields parameter $\Phi$ is defined as $u_\tau^2/u_g^2$,
where $u_\tau$ is the shear stress at the bottom.  Since the flow is similar to Poiseuille flow in
Case 1, the Shields parameter for Poiseuille flow $\Phi_{Pois}$ is computed from the fluid height
$h_f$:
\begin{equation}
  \Phi_{Pois} = \frac{6Re_{bulk}}{Ga^2}\left( \frac{d_p}{h_f} \right)^2.
  \label{eqn:shields}
\end{equation}
Additionally, the mean solid volume fraction of the bed is defined as:
\begin{equation}
  \varepsilon_{bed} = \frac{1}{y_2 - y_1} \int_{y_1}^{y_2}\varepsilon_s \mathrm{d}y,
  \label{eqn:epsilon_bed}
\end{equation}
where the interval is taken from $y_1 = 3d_p$ and $y_2 = 6d_p$ according to~\cite{kidanemariam14id}.

\subsection{Case 1: Flat Bed in Motion}
\label{sec:run1-flat}

The regime of the sediment pattern is determined by the Galileo number~\citep{charru06rf1}. At small
Galileo number (typically small particle or large viscosity) the flat bed is stable because the
erosion of the sediment bed is dominant. To demonstrate the capability of CFD--DEM in the prediction
of the `flat bed in motion' regime, numerical simulations are performed at relatively low Galileo
number. Detailed validation of the results obtained by using CFD--DEM is performed.

\begin{table}[!htbp]
  \caption{Parameters used in the simulations of `flat bed in motion'}
 \begin{center}
 \begin{tabular}{lcccccc}
   Case & $Re$ & $L_y$ & $h_f/d_p$ & $\Phi_{pois}$ & $\varepsilon_{bed}$ \\
   \hline
   Case 1a & 375  & 32 & 15.0 & 0.14 & 0.62 \\
   Case 1b & 750  & 32 & 15.1 & 0.27 & 0.62 \\
   Case 1c & 1125 & 32 & 14.6 & 0.43 & 0.61 \\
   Case 1d & 1500 & 32 & 14.0 & 0.63 & 0.58 \\
   Case 1e & 250  & 28 & 11.8 & 0.15 & 0.59 \\
   Case 1f & 500  & 28 & 11.2 & 0.33 & 0.63 \\
   Case 1g & 180  & 24 &  6.8 & 0.32 & 0.61 \\
   \hline
  \end{tabular}
 \end{center}
 \label{tab:param-flat}
\end{table}

Table~\ref{tab:param-flat} demonstrates the setup of the numerical simulations, which is based on
Case BL24 in the study by~\cite{kidanemariam14id} with variations in Reynolds number and channel
height.  Although the physical setup of Case 1a is the same as Case BL24, the fluid height $h_f$ in
the present simulation is slightly larger. This is because the interface-resolved model used
`contact length' and enlarged the distance between the sediment particles. Using the `contact
length' accounts for the compactness of the seabed but under-predicts the solid volume fraction
$\varepsilon_s$. Hence, this is not applied in the present simulation.

\begin{figure}[htbp]
  \centering
    \subfloat[]{
      \label{fig:flat-rate1}
      \includegraphics[width=0.45\textwidth]{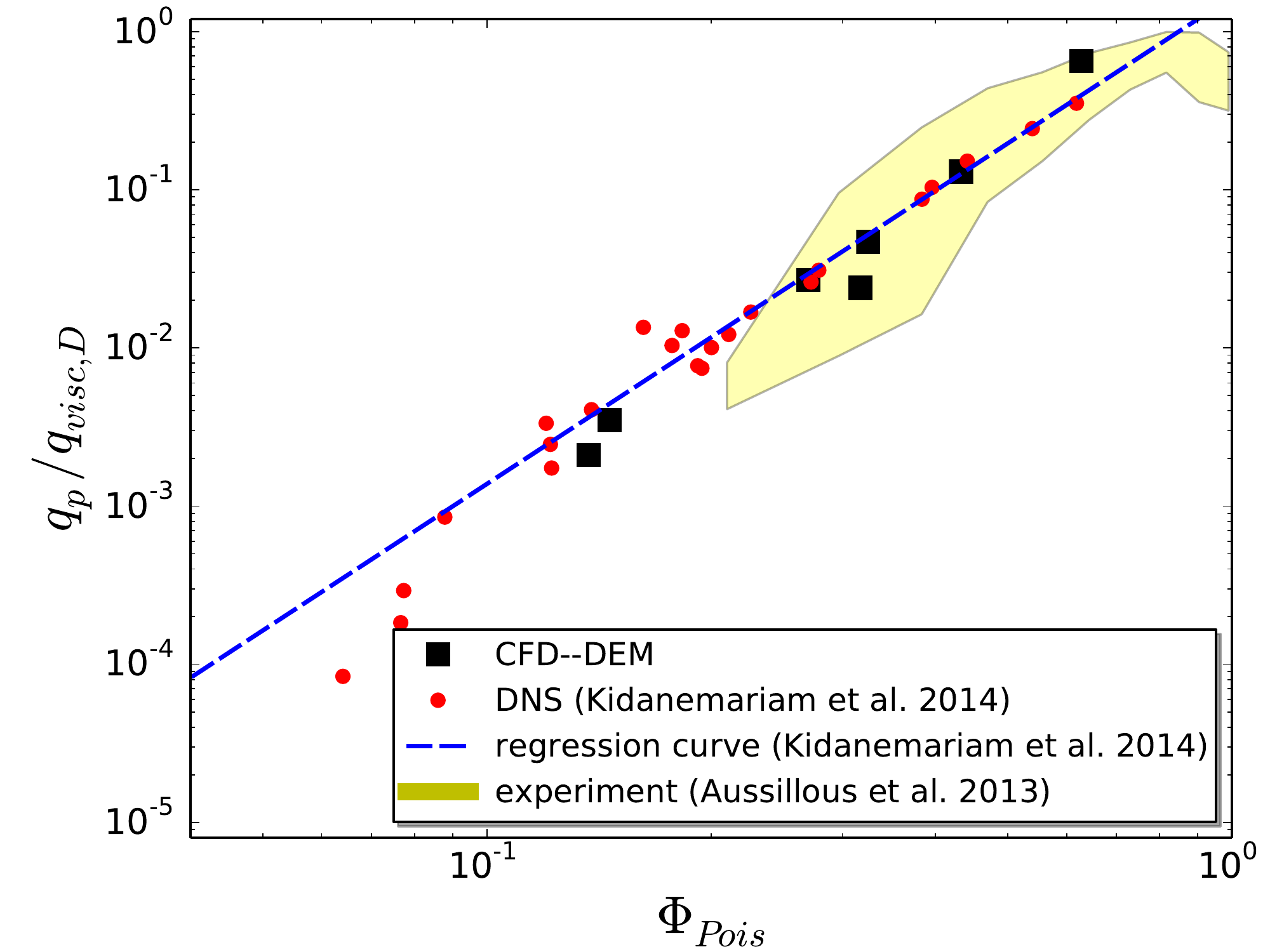}
    }
    \subfloat[]{
      \label{fig:flat-rate2}
      \includegraphics[width=0.45\textwidth]{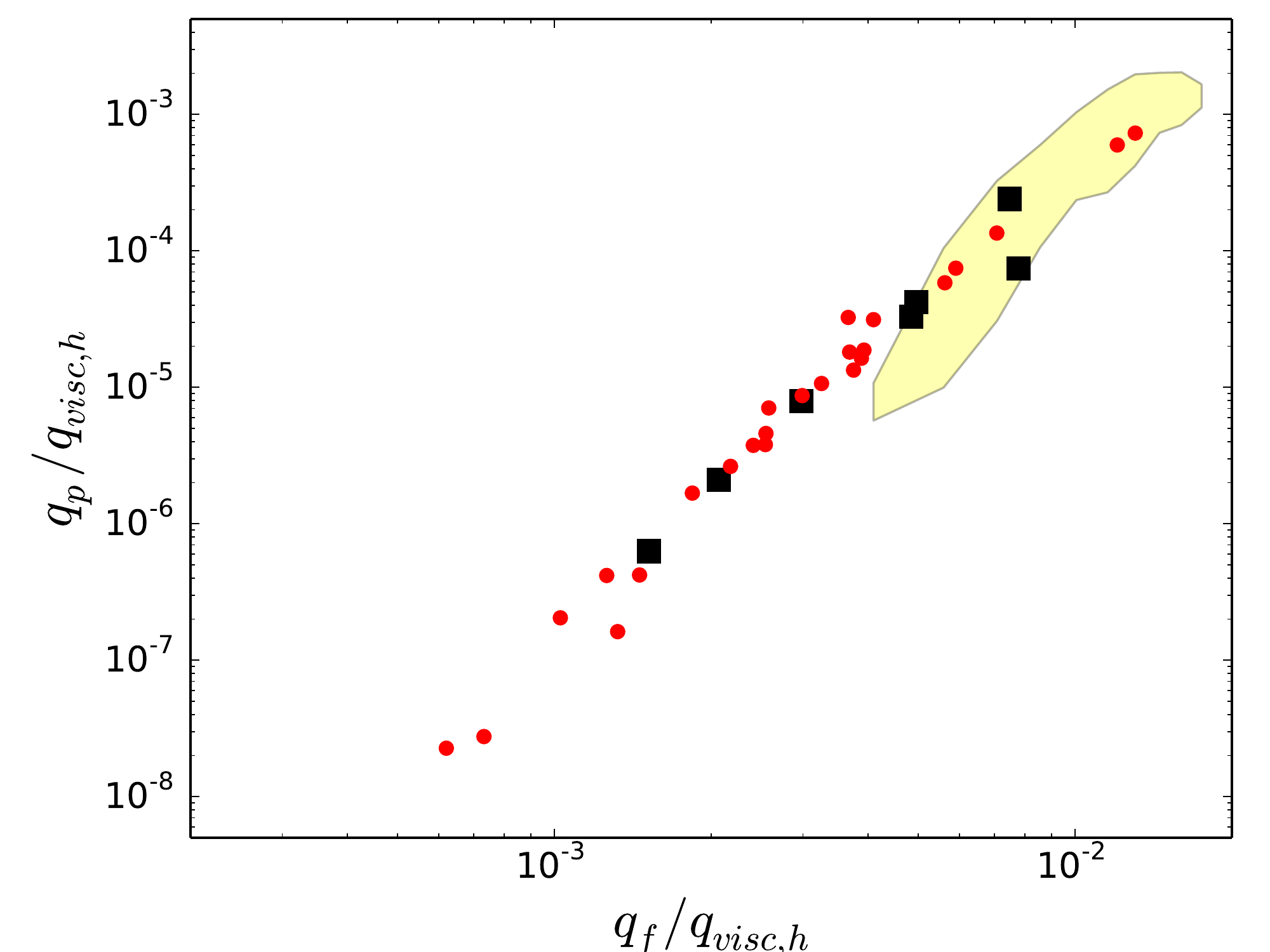}
    }
  \caption{Sediment transport rate obtained in the present simulations. (a) The sediment transport
    rate normalized by the viscous scaling $q_{visc,D}$ and plotted as a function of $\Phi_{Pois}$;
    (b) the sediment transport rate normalized by $q_{visc,h}$ and plotted as a function of fluid
    flow rate $q_f/q_{visc,h}$.}
  \label{fig:flat-sedi-rate}
\end{figure}

\begin{figure}[htbp]
  \centering
    \subfloat[]{
      \label{fig:flat-height1}
      \includegraphics[width=0.45\textwidth]{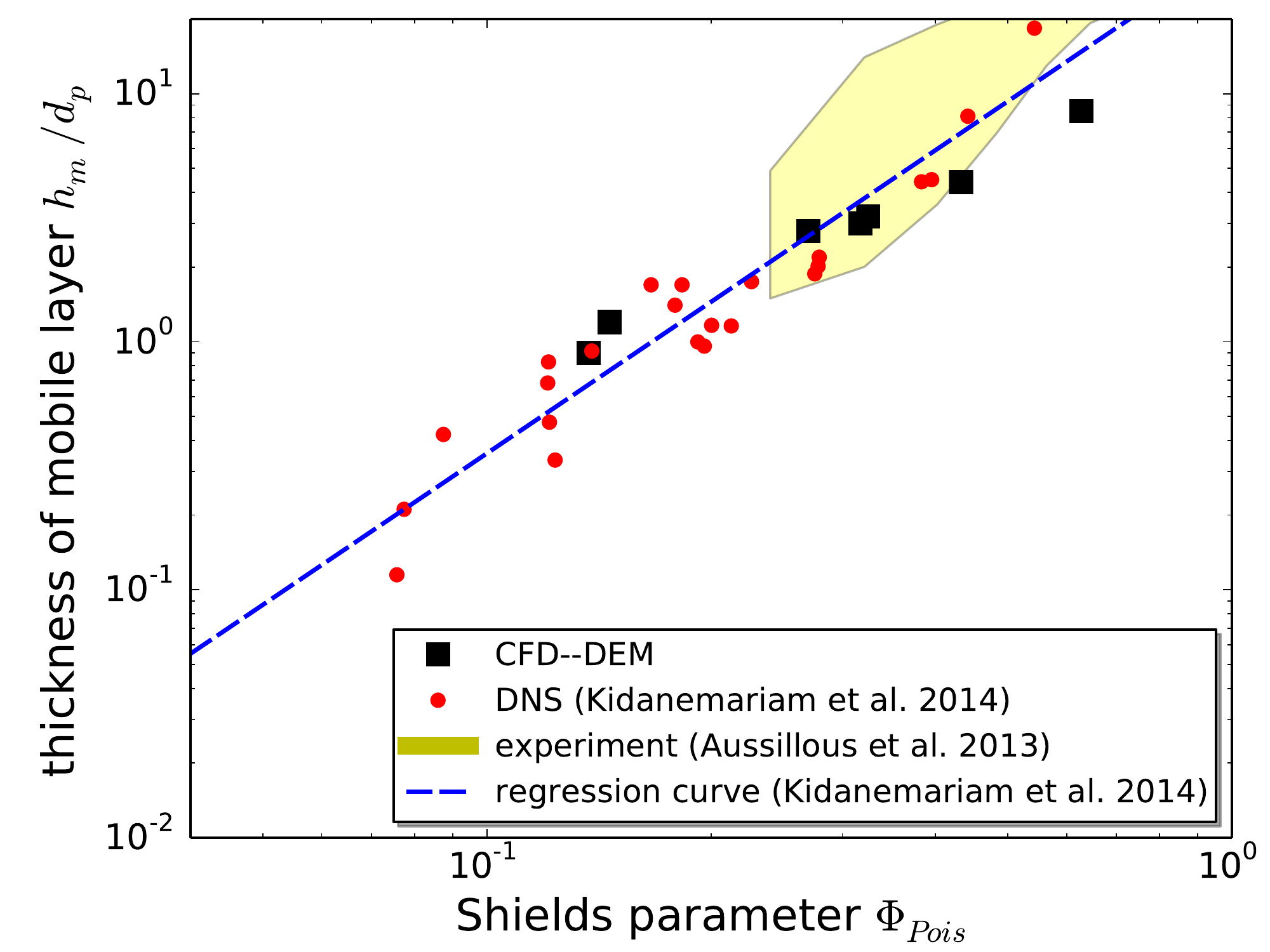}
    }
    \subfloat[]{
      \label{fig:flat-height2}
      \includegraphics[width=0.45\textwidth]{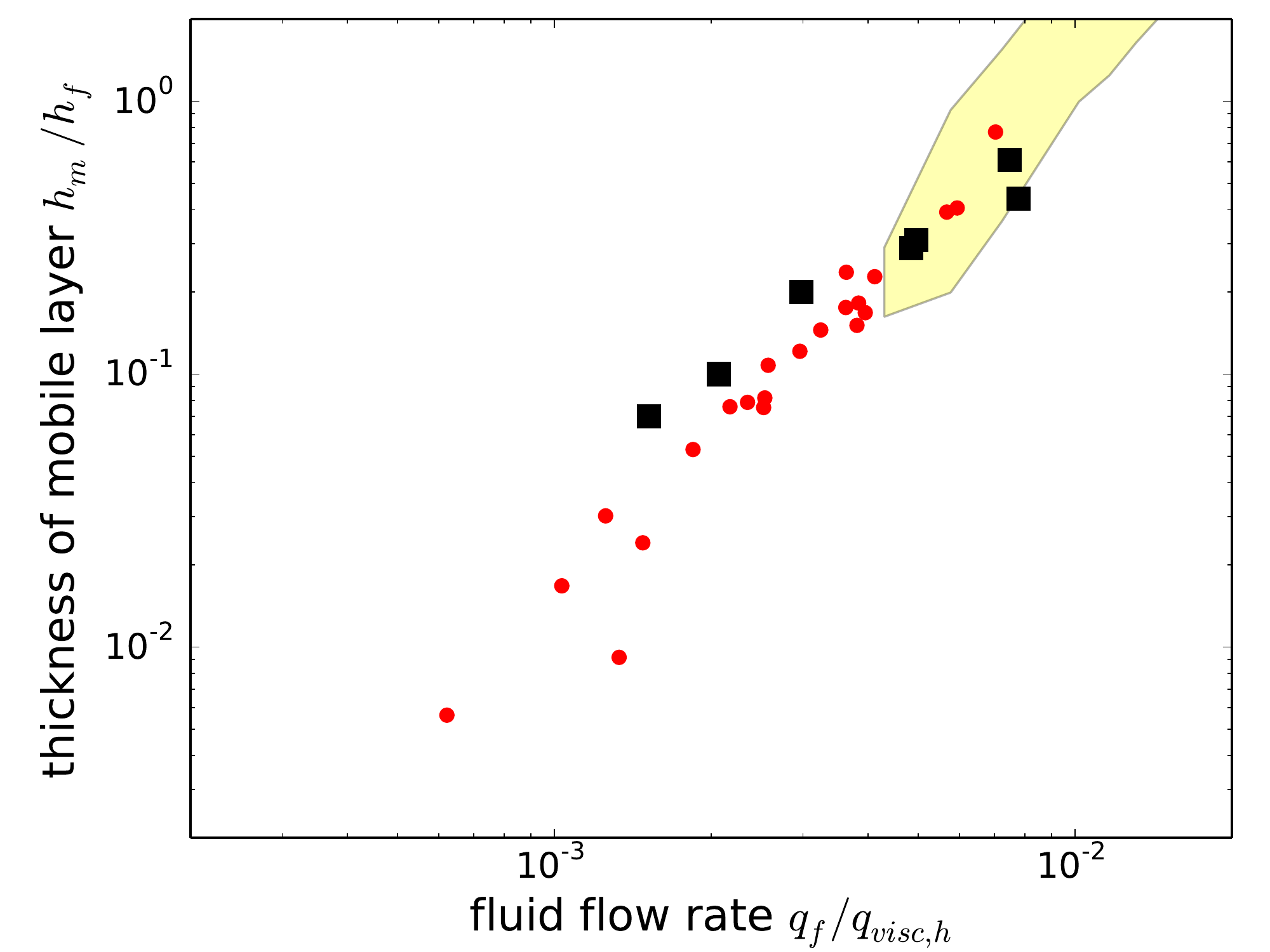}
    }
  \caption{The thickness of mobile layer obtained in the present simulations. (a) The height
    normalized by the particle diameter $d_p$ and plotted as a function of $\Phi_{Pois}$; (b) the
    height normalized by $h_f$ and plotted as a function of fluid flow rate $q_f/q_{visc,h}$.}
  \label{fig:flat-sedi-height}
\end{figure}

The sediment flux denotes the average velocity of the sediment particles, which is calculated as
follows:
\begin{equation}
  q_p(t) = \frac{\pi d_p^3}{6L_x L_z}u_{ave,p}(t),
  \label{eqn:sedi-flux-eqn}
\end{equation}
where $q_p(t)$ is the instantaneous sediment flux; $u_{ave,p}$ is the averaged particle velocity. To
normalize the sediment flux, the reference quantity $q_{visc,D}$ is used~\citep{aussillous13io}:
\begin{equation}
  q_{visc,D} = \frac{(\rho_p/\rho_f-1)g d_p^3}{\nu} = Ga^2\nu.
  \label{eqn:sedi-flux-norm1}
\end{equation}
Since $q_{visc,D}$ is based on the particle diameter $d_p$, another quantity $q_{visc,h}$ accounting
for the scale of fluid flow $h_f$ is used:
\begin{equation}
  q_{visc,h} = \frac{(\rho_p/\rho_f-1)g h_f^3}{\nu} = q_{visc,D}\left(\frac{h_f}{d_p}\right)^3.
  \label{eqn:sedi-flux-norm2}
\end{equation}
The time-averaged sediment flux $q_p$ obtained in the present simulations is shown in
Fig.~\ref{fig:flat-sedi-rate} to validate the CFD--DEM model. It can be seen that the sediment flux
$q_p$ normalized by both $q_{visc,D}$ and $q_{visc,h}$ are consistent with the trend of the data in
the literature. In addition, the sediment flux $q_p$ predicted by CFD--DEM is consistent with the
regression curve proposed by~\cite{kidanemariam14id}, in which the sediment flux increases cubically
($q_p/q_{visc,D} = 1.66 \Phi_{Pois}^{3.08}$). {\color{black} From the experimental
results~\citep{aussillous13io}, the critical Shields parameter is $0.12 \pm 0.03$. However, in both
present study and DNS simulations, the critical Shields parameter is not captured and the sediment
transport rate does not decrease significantly below the critical value. This may be attributed to
the fact that critical Shields parameter is defined by visual observation.} The averaged height of
moving particles $h_m$ is also demonstrated to evaluate the prediction of streamwise particle motion
under the particle-fluid interface. The definition of the moving particles uses the threshold value
$0.005 U_f^m$, where $U_f^m$ is the maximum flow velocity.  {\color{black} It can be seen in
Fig.~\ref{fig:flat-sedi-height} that the averaged height of the moving particles predicted using
CFD--DEM is consistent with the results obtained in DNS. The trend of the averaged height $h_m$ of
moving particles obtained in the present simulations is also consistent with the regression curve
obtained by~\cite{kidanemariam14id}, in which the height $h_m$ is proportional to the square of
Shields parameter ($h_m/d_p = 38.01 \Phi_{Pois}^{2.03}$). The height of the moving particles is of
the order of $10d_p$ in Case 1d when the Shields number is high.}

\begin{figure}[htbp]
  \centering
    \subfloat[]{
      \label{fig:flat-UMean}
      \includegraphics[width=0.45\textwidth]{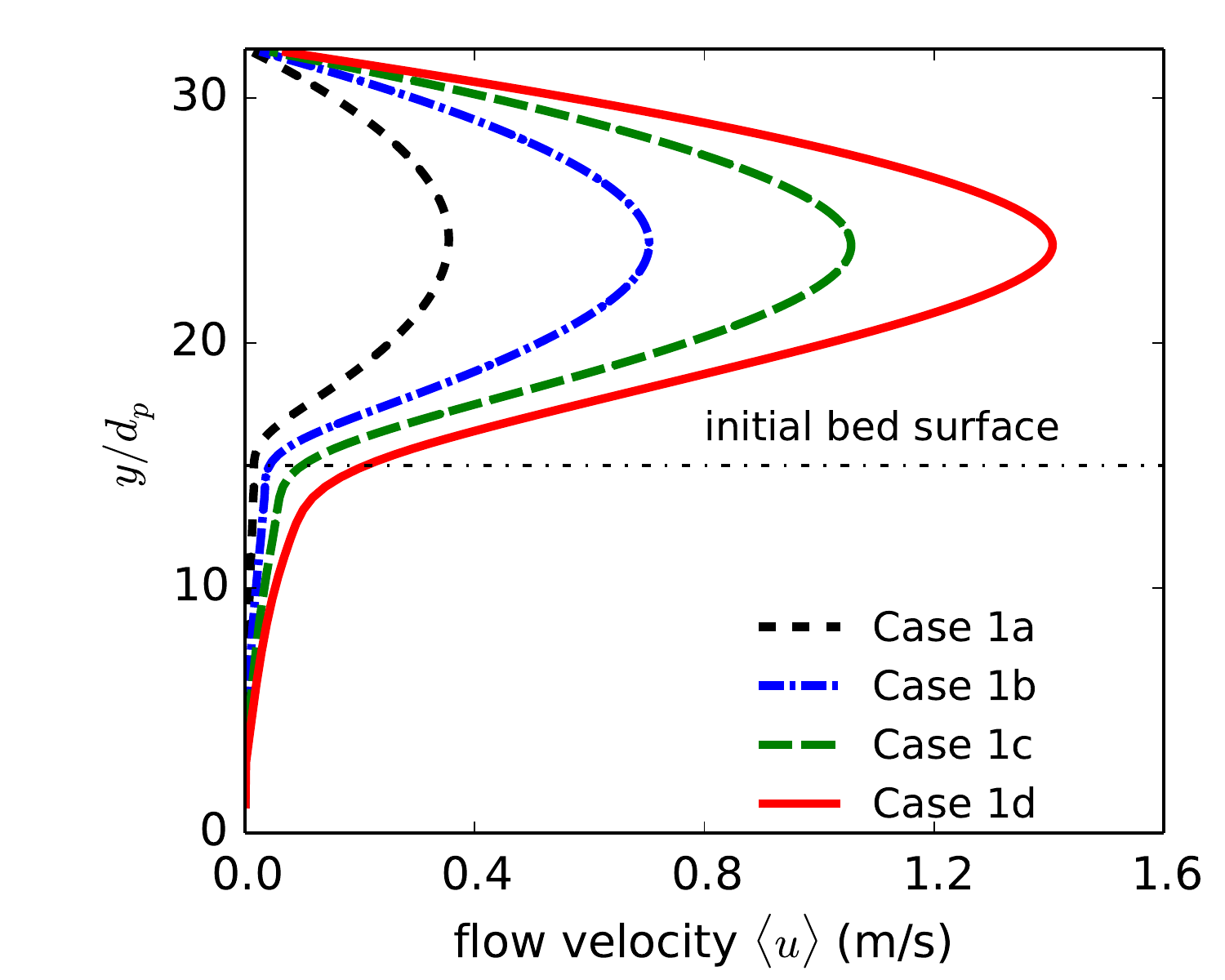}
    }
    \subfloat[]{
      \label{fig:flat-U-norm}
      \includegraphics[width=0.45\textwidth]{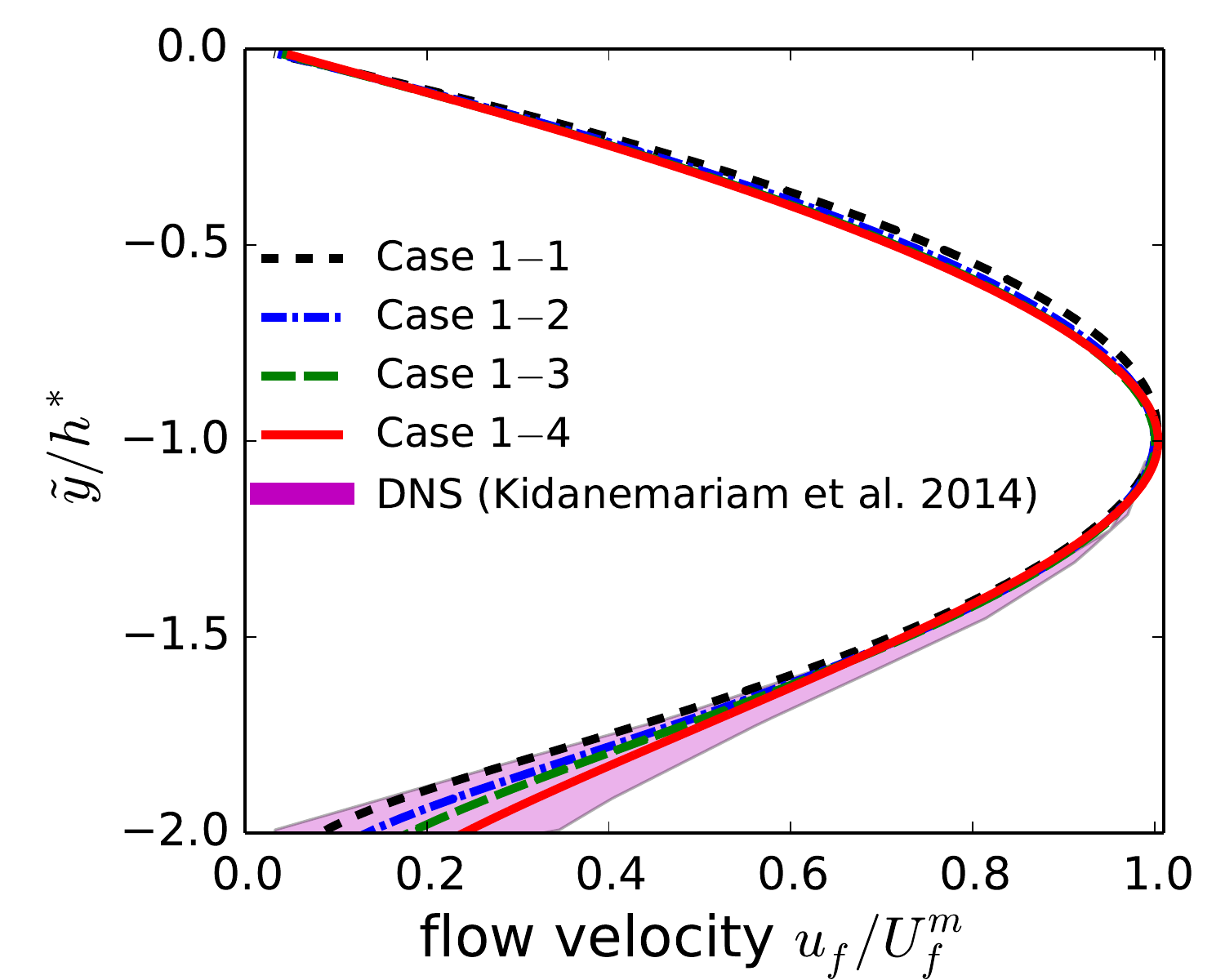}
    }
    \vspace{0.1in}
    \subfloat[]{
      \label{fig:UfmHf}
      \includegraphics[width=0.45\textwidth]{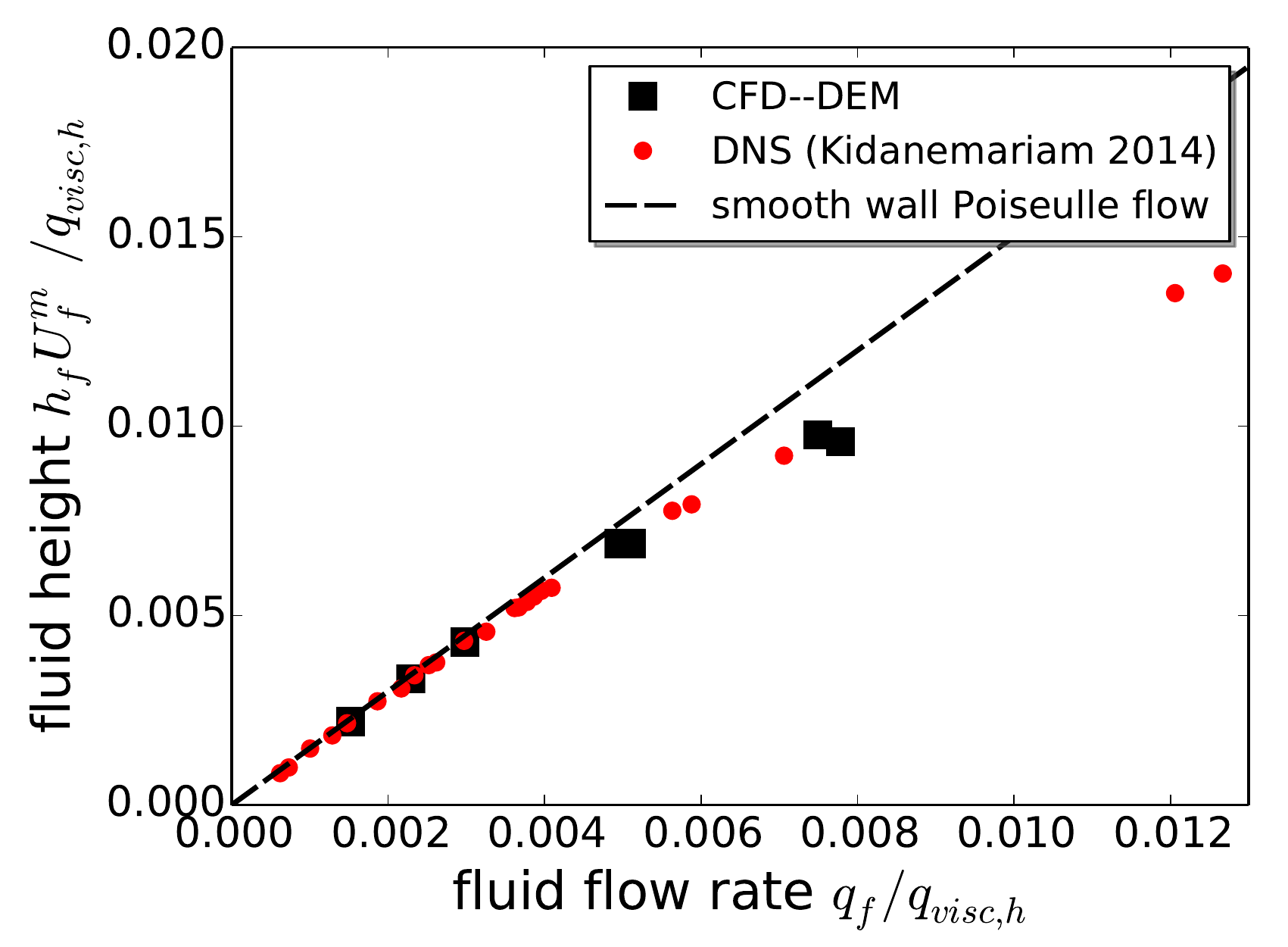}
    }
    \subfloat[]{
      \label{fig:h_star}
      \includegraphics[width=0.45\textwidth]{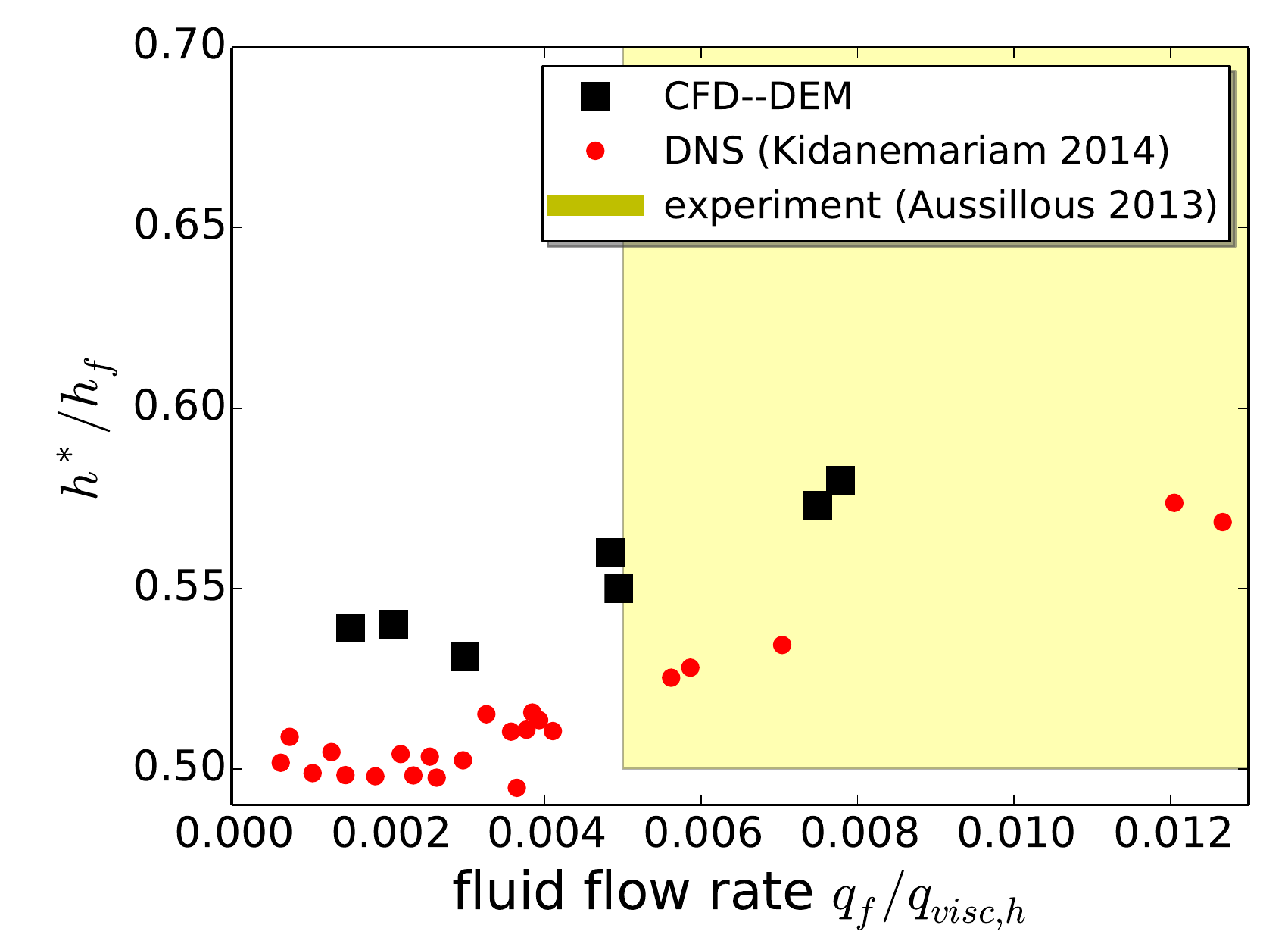}
    }
  \caption{The flow velocity and fluid height obtained in `flat bed in motion' regime. (a) The
    time-averaged flow velocity profile in the channel; (b) the normalized flow velocity profile by
    $U^m_f$; (c) the normalized fluid height plotted as a function of fluid flow rate
    $q_f/q_{visc,h}$; (d) the distance $h*$ between the location of $U^m_f$ and the top wall versus
    the normalized fluid flow rate.}
  \label{fig:flat-flow-all}
\end{figure}

Another quantity of interest in sediment transport is the velocity of the fluid flow. The flow
velocity profiles of Case 1a to Case 1d are shown in Fig.~\ref{fig:flat-flow-all}(a). The
normalized data are plotted in Fig.~\ref{fig:flat-flow-all}(b): the flow velocity is normalized by
the maximum flow velocity $U_f^m$, and the distance to the top wall $\tilde{y}=y - L_y$ is
normalized by $h^*$, which is the distance between the top wall and the location of $U_f^m$. It can
be seen in Fig.~\ref{fig:flat-flow-all}(b) that the flow velocity profiles predicted by using
CFD--DEM are consistent with the results obtained using interface-resolved method. It is noted that
the flow under the fluid-sediment interface is nonzero, which is consistent with the experimental
observations~\citep{aussillous13io}. This is an improvement of the DEM-based method over
hydro-morphodynamics models, since the hydro-morphodynamics models cannot capture the flow velocity
under sediment bed~\citep{hydraulics99df,nabi13ds3}.  Fig.~\ref{fig:flat-flow-all}(c) demonstrates
the relationship of the normalized fluid height $h_f U_f^m/q_{visc,h}$ and the fluid flux
$q_f/q_{visc,h}$. At small fluid flux, a linear relationship is observed, which indicates the flow
is similar to Poiseuille flow; at relatively large fluid flux the CFD--DEM model captures the
deviation of the maximum fluid velocity from Poiseuille flow. The distance $h^*$ between the top
wall and maximum flow velocity $U_f^m$ can be used to evaluate the influence of sediment bed to the
fluid flow. The scattered plot of $h^*$ as a function of the fluid flux in dimensionless form is
shown in Fig.~\ref{fig:flat-flow-all}(d). It can be seen that the prediction of $h^*$ are consistent
with the experimental data ranging from 0.5-0.7.

\begin{figure}[!htpb]
 \centering
  \subfloat[$t_0$]{
      \includegraphics[width=0.45\textwidth]{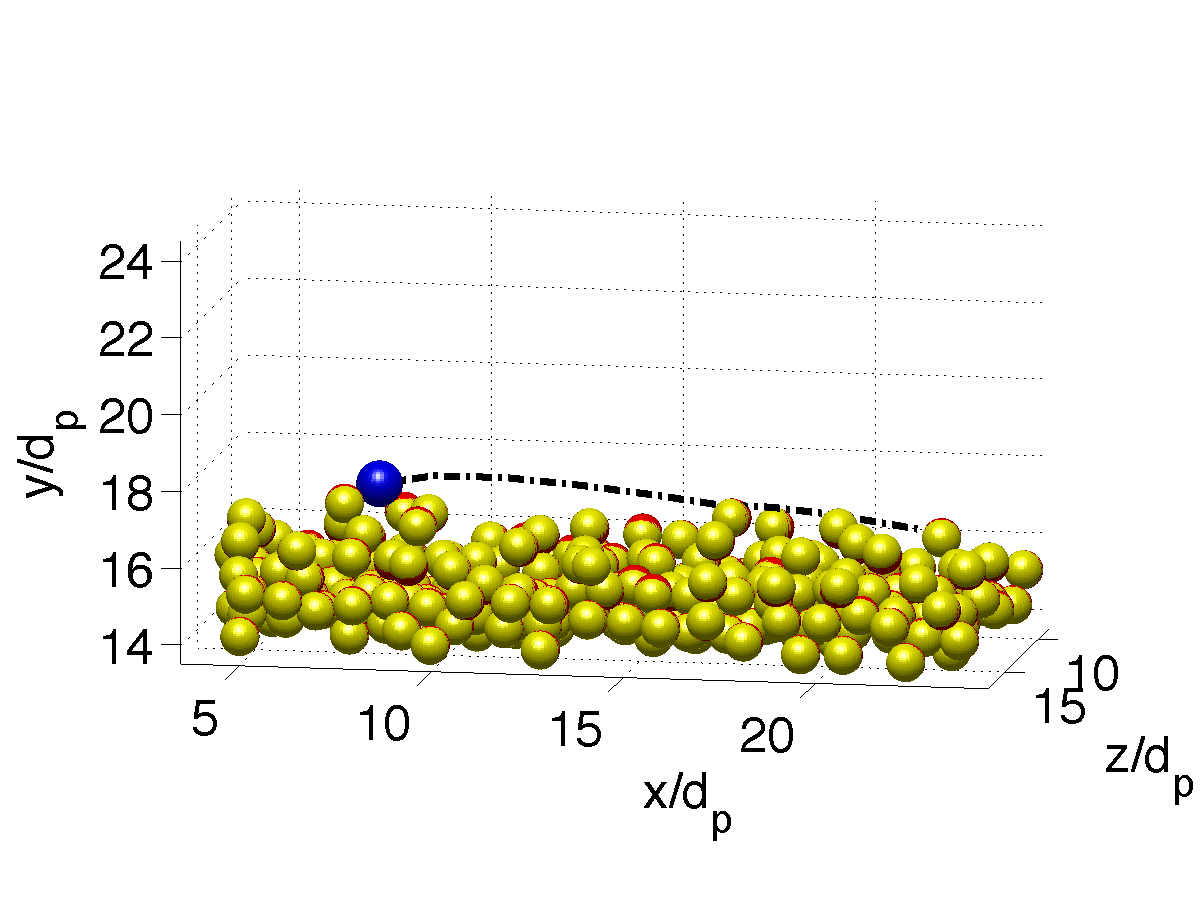}
  }
  \subfloat[$t_0 + \Delta t$]{
      \includegraphics[width=0.45\textwidth]{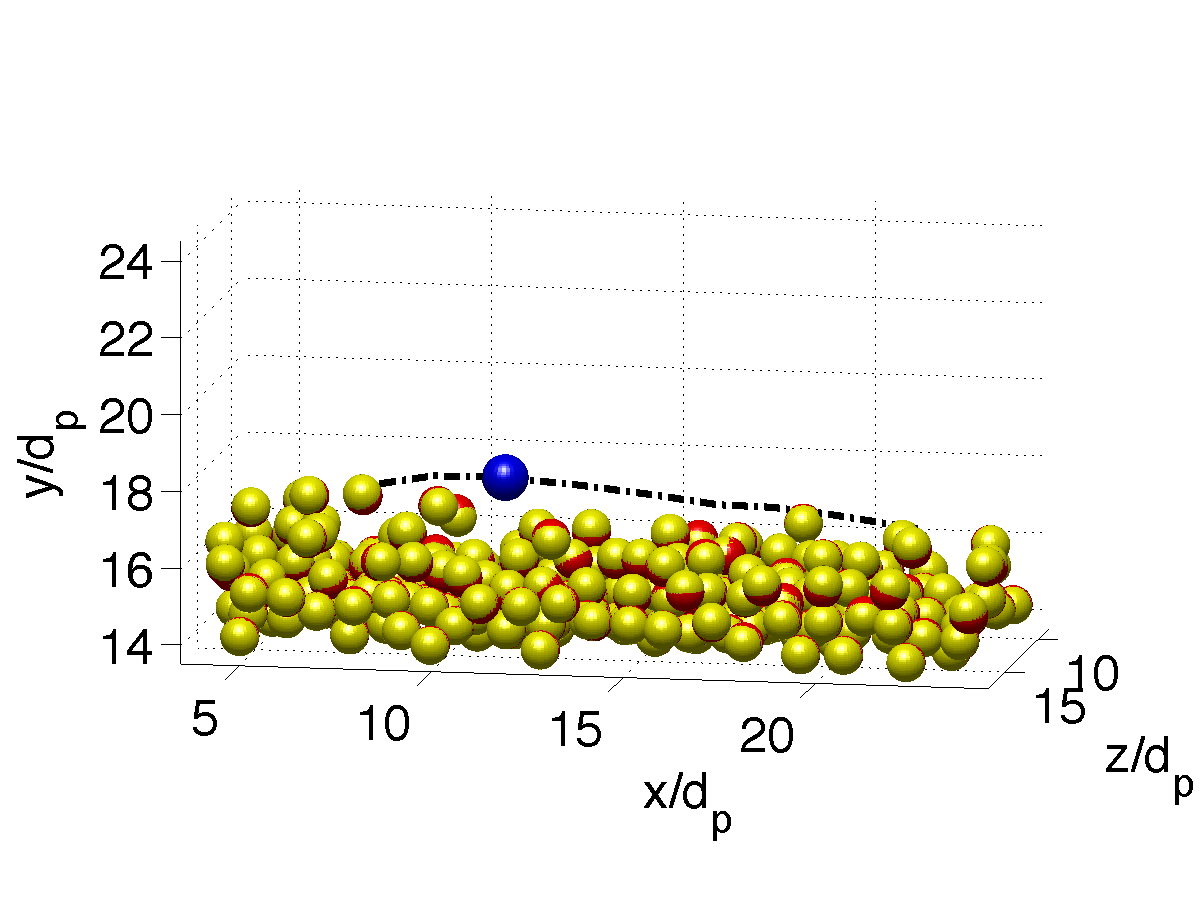}
  }
  \hspace{0.001\textwidth}
  \subfloat[$t_0 + 2\Delta t$]{
      \includegraphics[width=0.45\textwidth]{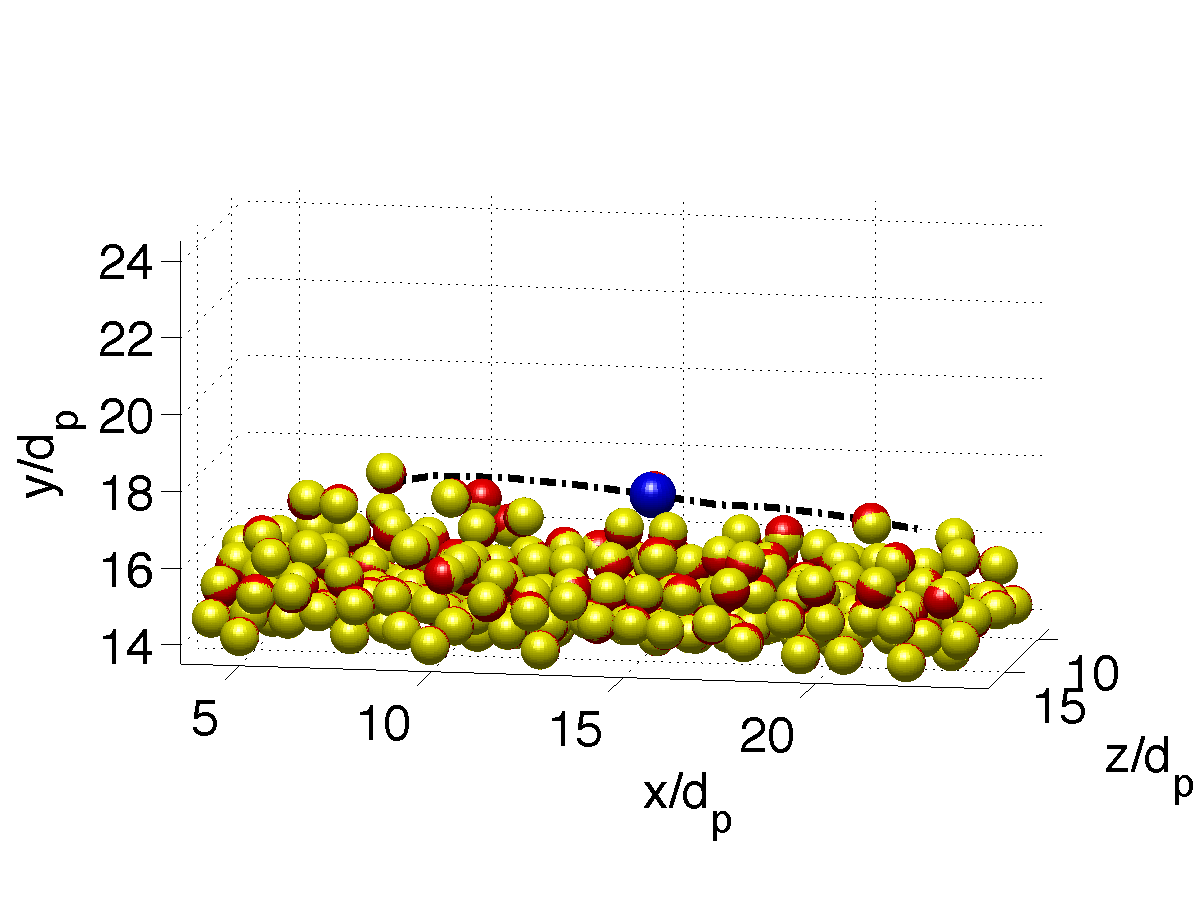}
  }
  \subfloat[$t_0 + 3\Delta t$]{
      \includegraphics[width=0.45\textwidth]{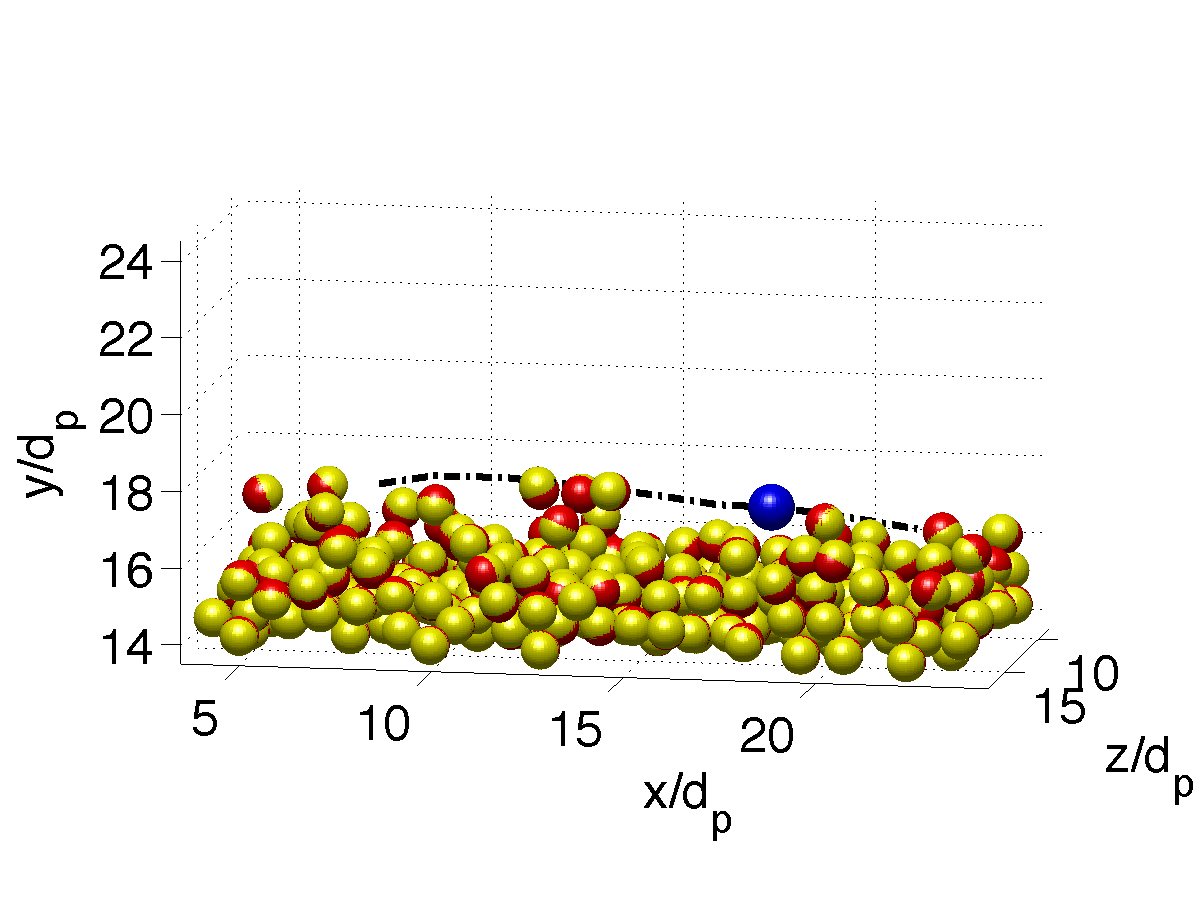}
  }
 \caption{Snapshots of particle locations during the saltation of the highlighted particle (colored
   in blue). To avoid cluttering, only a portion at the top of sediment bed is shown. The red
   dash-dot line indicates the trace of particle saltation of the highlighted particle. The
   particles are colored half yellow and half red to indicate the rotation. Only the particles on
   the surface are plotted. The time origin ($t_0$) corresponds to the beginning of the cycle and
   the time interval $\Delta t = 0.45 tU^m_f/L_x$, which corresponds to 0.45 flow-through times or
   1000 DEM time steps. The x,y,z axis are normalized by using the particle diameter. Animations of
   the particle trajectories of are provided in the supplementary materials.
 }
 \label{fig:salt-ani}
 \end{figure}

{\color{black}
To illustrate the motion of the sediment particles in `flat bed in motion' regime, the snapshots of
the motion of sediment particles in Case 1b are shown in Fig.~\ref{fig:salt-ani}. The snapshots at
$t_0$ correspond to the start of the saltation and the time interval between the snapshots $\Delta t
= 0.45 tU^m_f/L_x$, which corresponds to 0.45 flow-through times or 1000 DEM time steps. The sliding
motion of the particles can be observed according to the change of the locations in the snapshots.
To display the particle rotation, the sediment particles are colored half yellow and half red. The
yellow halves of the particles are in front at time~$t_0$, and rotation of the sediment particles
can be seen from the change of the orientations. In addition, the saltation of the particle
(highlighted in blue) is also shown in Fig.~\ref{fig:salt-ani}. The trace of the blue particle is
plotted in the snapshots using the black dash-dot line. It can be seen that the highlighted particle
jumps a distance of approximately 10 diameters along the sediment bed at the height of one particle
diameter. From the results shown in the snapshots, it can be seen that the sliding, rotation and
saltation of sediment particles in bedload are captured by CFD--DEM simulations. Animations of the
particle trajectories of Case 1b are provided in the supplementary materials.
}

\subsection{Case 2: Generation of Dunes}
\label{sec:run2-dune}

CFD--DEM is used to study the evolution of different dunes according to the regime map in
Fig.~\ref{fig:dune-regime}. This is to demonstrate the capability of CFD--DEM in the prediction of
dune migration. It can be seen from the regime map that the dune height increases with Galileo
number, which is due to the increase of particle inertia. Simulations at different Galileo numbers
are performed to show that CFD--DEM is able to predict the generation of both `small dune' and
`vortex dune'.  The results obtained by using CFD--DEM are validated using numerical benchmark and
experimental data.

\begin{figure}[htbp]
  \centering
    \subfloat[small dune]{
      \label{fig:dune1-his}
      \includegraphics[width=0.45\textwidth]{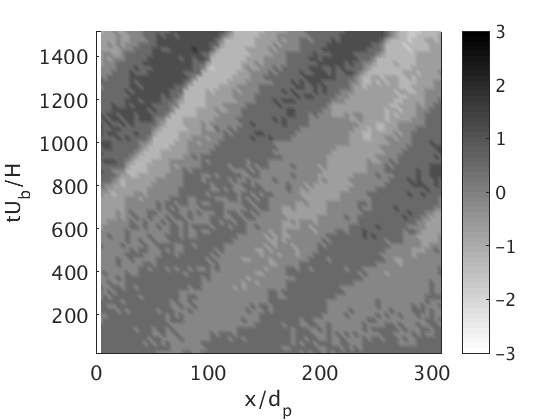}
    }
    \subfloat[vortex dune]{
      \label{fig:dune2-his}
      \includegraphics[width=0.45\textwidth]{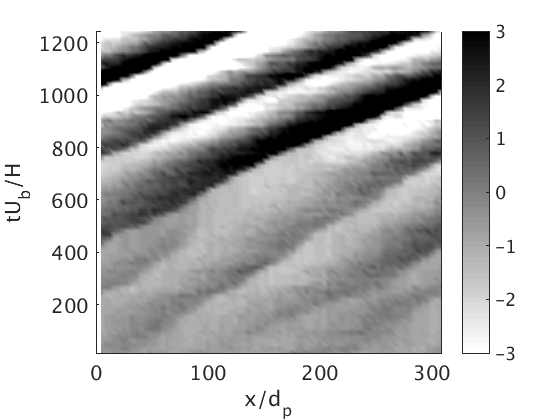}
    }
  \caption{Space-time revolution of the fluid-bed interface $h' = h(t) - \bar h(t)$ for (a) small
  dune and (b) vortex dune, normalized by the particle diameter $d_p$.}
  \label{fig:dune-his}
\end{figure}

\begin{figure}[htbp]
  \centering
    \subfloat[small dune]{
      \label{fig:dune1-final}
      \includegraphics[width=0.9\textwidth]{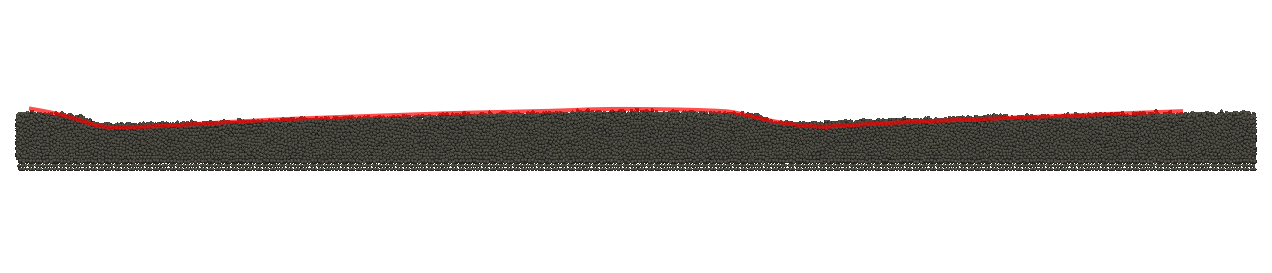}
    }
    \vspace{0.1in}
    \subfloat[vortex dune]{
      \label{fig:dune2-final}
      \includegraphics[width=0.9\textwidth]{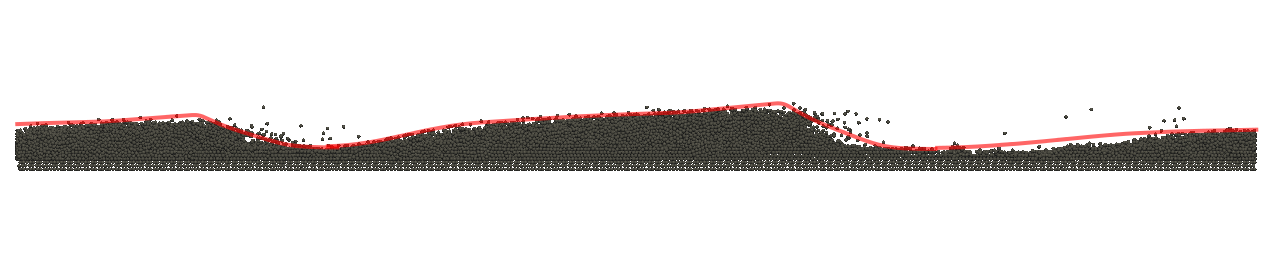}
    }
  \caption{Typical snapshots of the dune shape obtained in present simulations. The black particles
  denotes the DEM particles; the red curves are geometrically similar to the dunes surfaces obtained
  from the experimental measurements~\citep{ouriemi09sd2}.}
  \label{fig:dune-final}
\end{figure}

The space-time evolutions of both `small dune' and `vortex dune' are shown in
Fig.~\ref{fig:dune-his}. In this figure, the evolution of sediment patterns obtained by using
CFD--DEM are similar with those observed in the study by~\cite{kidanemariam14dn}. The dunes are
developed from the perturbations on the flat bed, which is consistent with the observations
by~\cite{kidanemariam14dn}. The dune height, wavelength, and the migration velocity of the dunes
obtained in the present simulations are demonstrated in Table~\ref{tab:dune-tab}. It can be seen
that the accuracy of CFD--DEM is satisfactory compared with the numerical benchmark and experimental
data. Note that the dune migration velocity predicted by using CFD--DEM is smaller than the result
obtained by using interface-resolved model. This is because the compactness of the sediment bed in
the interface-resolved model is smaller when using the 'contact length', and thus the sediment
particles are more likely to become suspended and move faster. Since the particles on the crest move
more rapidly, the migration velocity of the dune is larger. The snapshot of the dune shape obtained
in the present simulations is shown in Fig.~\ref{fig:dune-final}.  It can be seen that both the
`small dune' and `vortex dune' generated by using CFD--DEM are geometrically similar with the
experimental data obtained by~\cite{ouriemi09sd2}.

\begin{figure}[htbp]
  \centering
    \subfloat[$tU_b/H = 560$]{
      \label{fig:dune-Q1}
      \includegraphics[width=0.6\textwidth]{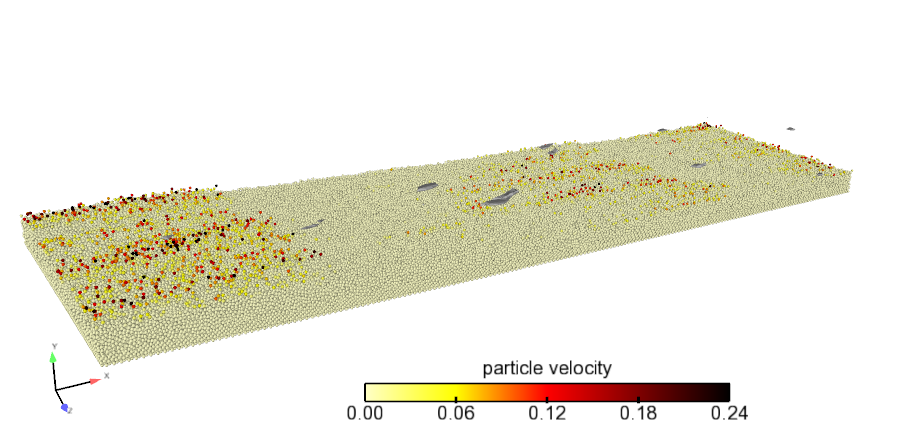}
    }
    \vspace{0.1in}
    \subfloat[$tU_b/H = 720$]{
      \label{fig:dune-Q2}
      \includegraphics[width=0.6\textwidth]{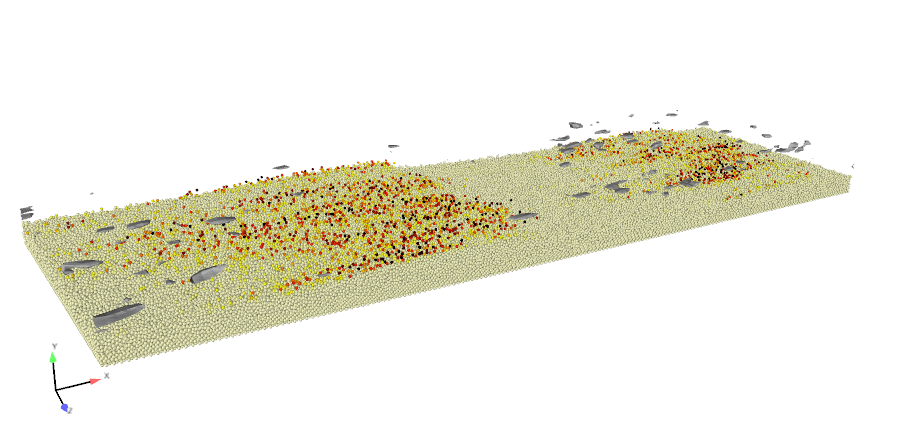}
    }
    \vspace{0.1in}
    \subfloat[$tU_b/H = 880$]{
      \label{fig:dune-Q3}
      \includegraphics[width=0.6\textwidth]{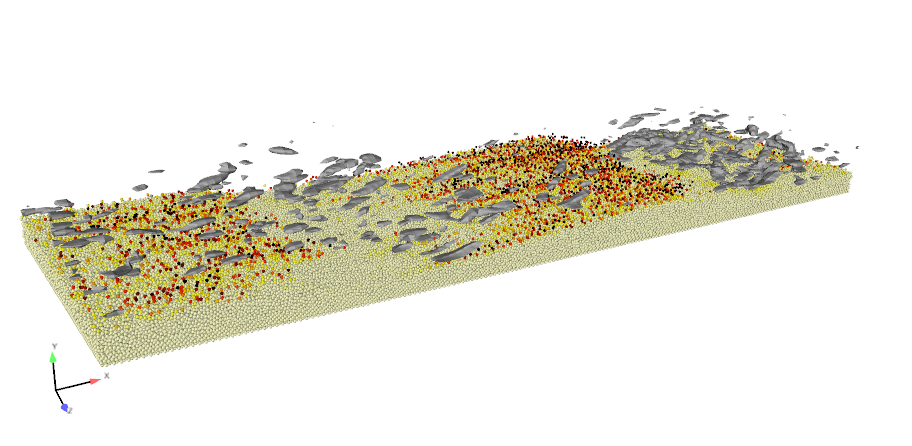}
    }
  \caption{Vortice after the vortex dune using Q-criterion at three snapshots. The iso-surface of $Q
  = 2000$ is plotted, which is the second invariant of the velocity gradient tensor. The unit of the
  particle velocity is m/s. The development of the vortical structures is available in supplementary
  materials.}
  \label{fig:dune-Q}
\end{figure}

The iso-surface of Q-criterion for Case 2b is shown in Fig.~\ref{fig:dune-Q}. It can be seen that
CFD--DEM captures the vortex shedding after the dune crests. The vortical structure obtained in the
present simulation is consistent with the predictions by~\cite{zedler01ls}.  Additionally, the
comparison of Figs.~\ref{fig:dune-Q}(a), \ref{fig:dune-Q}(b) and \ref{fig:dune-Q}(c)
demonstrates that the number of vortices increases with the height of the dune.  This is
consistent with the conclusion in the literature that vortical structure is significantly influenced
by the dunes~\citep{zedler01ls,nabi13ds3,arolla15tm}. 


\begin{table}[!htbp]
  \caption{Comparison of the dune properties in different tests.}
 \begin{center}
 \begin{tabular}{lccc}
   Small dune               & present simulation & interface-resolved model & experimental results\\
   \hline
   dune height              & $ 2d_p$   &   $ 2d_p$     & $ 2.5d_p$ \\
   wavelength               & $ 156d_p$ &   $ 140d_p$   & $ 130d_p$ \\
   migration velocity       & $ 0.004u_b$ & $ 0.011u_b$ & 0.0024--0.01$u_b$ \\
   \hline
   Vortex dune               & & &  \\
   \hline
   dune height              & $ 7d_p$ &   $ 5d_p$    & 4--8$d_p$ \\
   wavelength               & $ 156d_p$ &   $ 153.6d_p$  & $ 150d_p$ \\
   migration velocity       & $ 0.016u_b$ & $ 0.035u_b$  & 0.01--0.03$u_b$ \\
   \hline
  \end{tabular}
 \end{center}
 \label{tab:dune-tab}
\end{table} 

\begin{figure}[htbp]
  \centering
      \label{fig:dune-rate1}
      \includegraphics[width=0.45\textwidth]{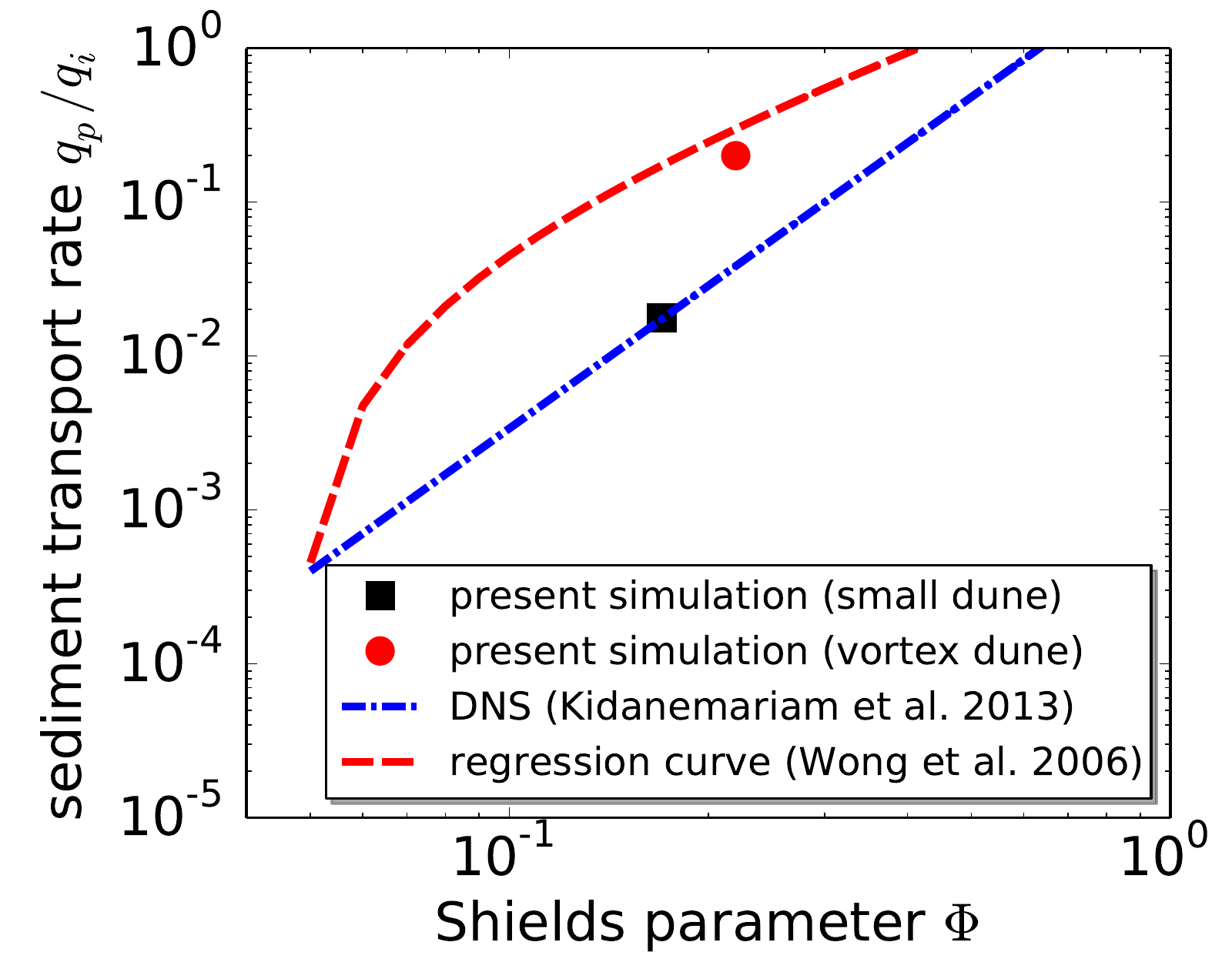}
  \caption{The sediment transport rate $q_p$, normalized by $q_i = u_g d_p$, of Case 2a and 2b
    plotted as a function of Shields parameter. The Shields parameter $\Phi_{pois}$ of Case 2a is
    defined using Eq.~(\ref{eqn:shields}); the Shields parameter of Case 2b is defined as $\Phi =
  u_\tau^2/u_g^2$.}
  \label{fig:dune-rate}
\end{figure}

The sediment transport rates from the present simulations are compared with the
experimental data in Fig.~\ref{fig:dune-rate}. It can be seen that the sediment transport rates of
both the `small dune' regime and `vortex dune' regime are consistent with the experimental
results~\citep{aussillous13io, wong06rc}. This agreement supports the conclusion that the dunes
formed at the bottom do not significantly influence the sediment transport
rate~\citep{kidanemariam14dn}. It is noted that the sediment transport rates obtained in Case 2b is
significantly larger than the predictions of Case 2a. This is because the flow regimes of the two
test cases are different. In the `small dune' generation test, the particles are rolling and sliding
on the sediment bed in laminar flow, and thus there is only bedload. In contrast, in the `vortex dune'
case, the flow is turbulent and thus the suspended load contributes to the total sediment
flux. In turbulent flow, the sediment particles move much faster than the particles rolling on the
sediment bed. Therefore, the sediment transport rate in turbulent flow is larger than that in
laminar flow even at the same Shields parameter.

In summary, despite some discrepancies, the overall agreement of the results obtained by using
\textit{SediFoam} and those in the literature is good. Compared with the computational
costs of interface-resolved method ($5\times10^6$ computational hours), the costs of the Case 1b
are only $2\times10^4$ computational hours, which is more than two orders of magnitude smaller. The
comparison between CFD--DEM and interface-resolved method suggests that CFD--DEM can predict the
movement of dune generation with satisfactory accuracy by using much smaller computational costs.

\subsection{Case 3: Suspended Particles}
\label{sec:run3-suspend}

In turbulent flow, if the vertical component of the eddy velocity is larger than the terminal
velocity of the sediment particle, the particles become suspended. Simulations are performed to
demonstrate the capability of CFD--DEM in `suspended load' regime. In addition, the improvements of
the present model over existing simulations using CFD--DEM are presented. The results obtained are
only validated using experimental results. This is because the simulations using interface-resolved
method are not available in the literature due to the computational costs for high Reynolds number
flows.

\begin{figure}[htbp]
  \centering
    \subfloat[Sediment transport rate]{
      \label{fig:sedi-rate}
      \includegraphics[width=0.45\textwidth]{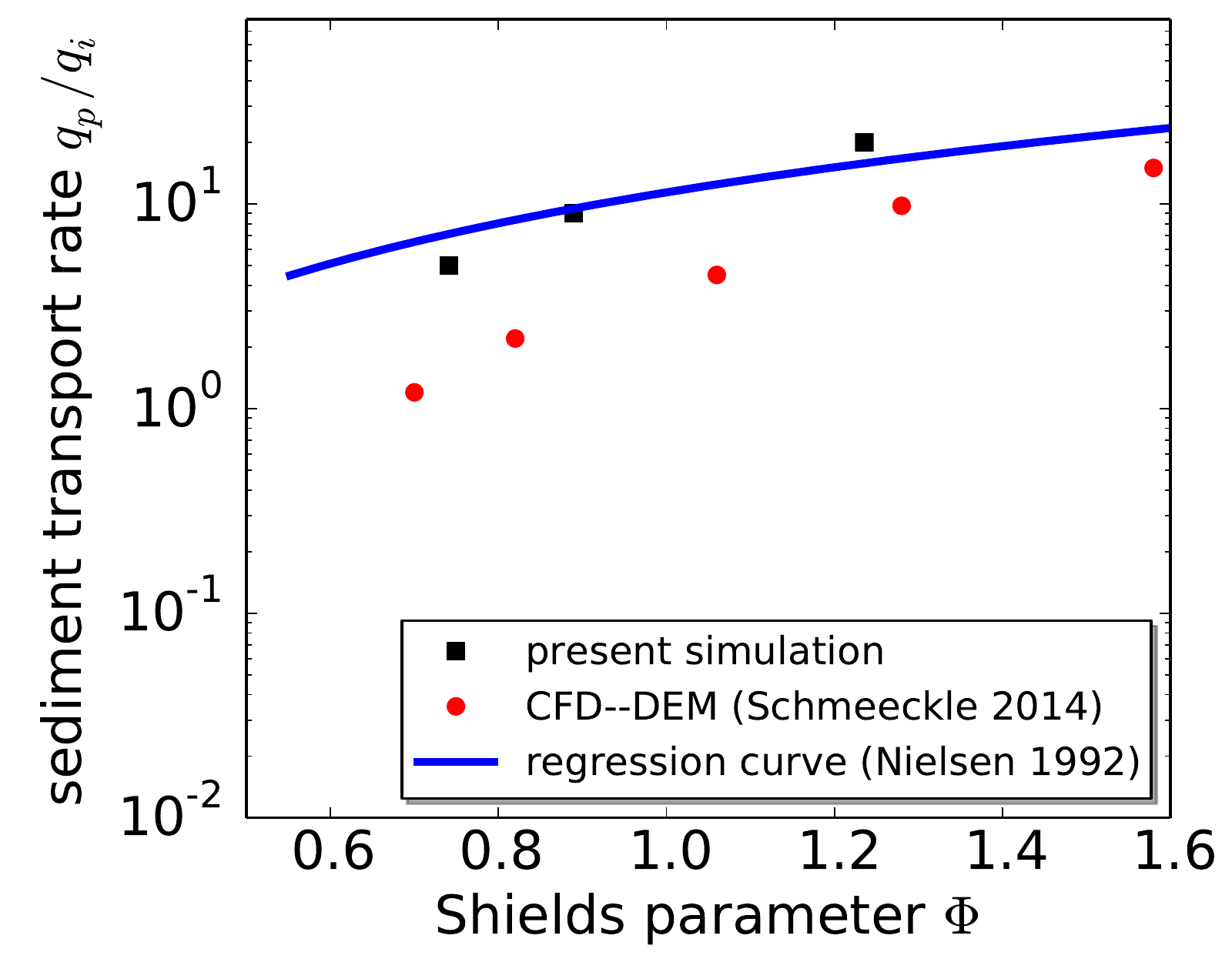}
    }
    \hspace{0.001\textwidth}
    \subfloat[Surface friction]{
      \label{fig:sedi-cf}
      \includegraphics[width=0.45\textwidth]{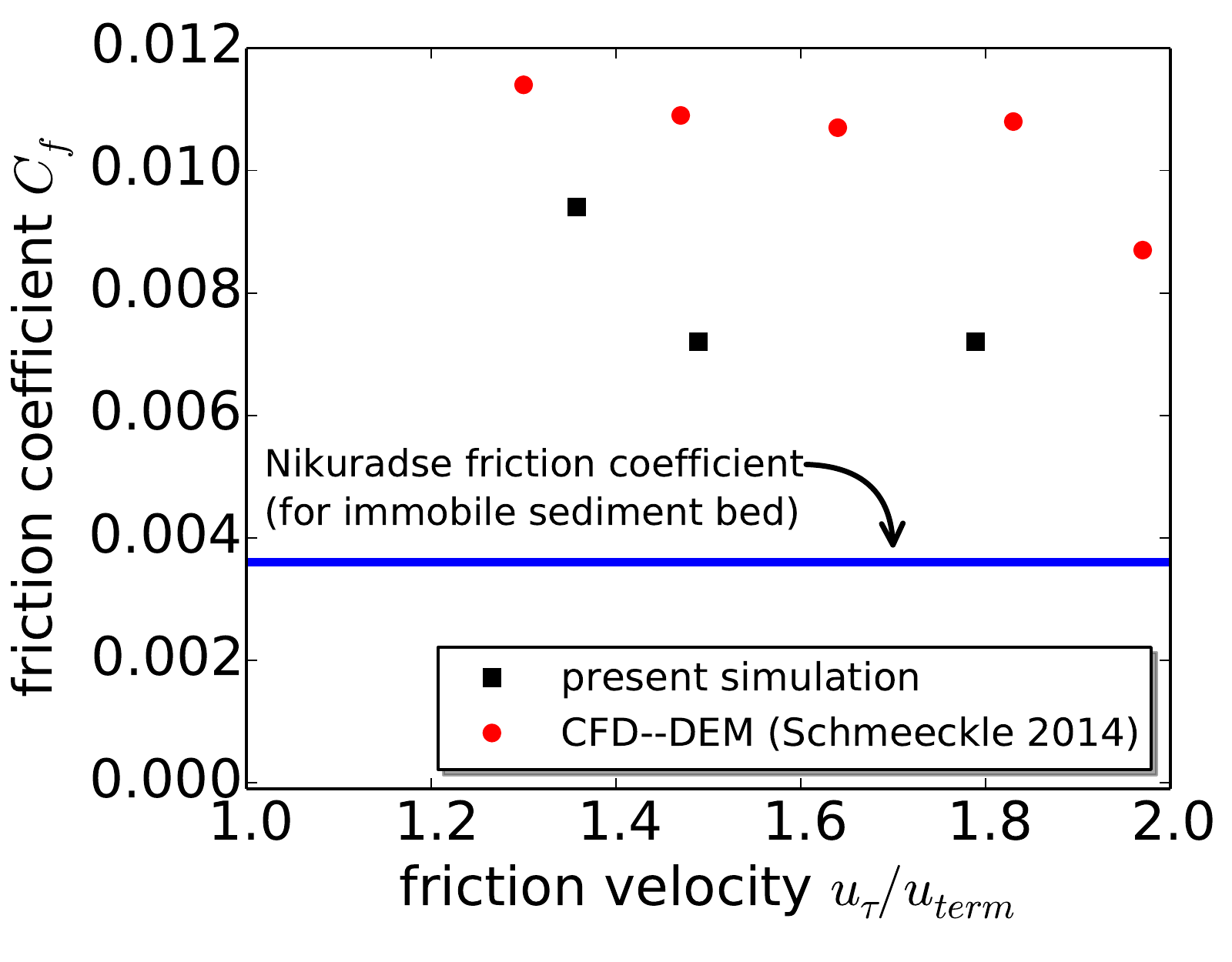}
    }
    \caption{The average properties of the sediment particles. (a) Sediment transport rate; (b)
    surface friction.}
  \label{fig:sedi2-rate}
\end{figure}

The domain geometry, the mesh resolution, and the properties of fluid and particles are detailed in
Table~\ref{tab:param-sedi}. The flow velocities in three numerical simulations range from $0.8$~m/s
to $1.2$~m/s. The averaged properties of sediment particles are presented in
Fig.~\ref{fig:sedi2-rate}, including the sediment transport rate and the friction coefficient. It
can be seen that the sediment transport rates agree favorably with the experimental
data~\citep{nielsen92cb}. {\color{black} It is worth mentioning that the prediction of sediment
transport rate $q_p/q_i$ using \textit{SediFoam} agrees better with the experimental data than the
results obtained by~\cite{schmeeckle14ns}. In the present simulations, the shielding effect of
particles is considered by accounting for solid volume fraction $\varepsilon_s$, and thus the
terminal velocity of sediment particles at the seabed is smaller. Since the terminal velocity of the
particles is smaller, the particles are more likely to move faster so that the predicted sediment
transport rate is larger.} The coefficient of friction of the surface is defined as:
\begin{equation}
  C_f = u_{\tau}^2/\langle u \rangle^2,
  \label{eqn:frictionCoef}
\end{equation}
which describes the hydraulic roughness. As shown in Fig.~\ref{fig:sedi2-rate}(b), $C_f$ obtained in
the present simulation and by~\cite{schmeeckle14ns} are larger than the Nikuradse value obtained by
using immobile seabed. The increase in $C_f$ is because the hydraulic roughness over a loose bed is
larger in the presence of movable particles~\citep{schmeeckle14ns}. {\color{black} Note that the
friction coefficient $C_f$ predicted by \textit{SediFoam} is slightly smaller than the results
predicted by~\cite{schmeeckle14ns}. In the present simulation, the volume averaged fluid velocity is
obtained by using $\langle u \rangle = \int\limits_{V} \varepsilon_f \mathbf{U}_{f,x}
\,dV/\int\limits_{V} \varepsilon_f \,dV$. When the volume fraction term is considered, the fluid
volume fraction $\varepsilon_f < 1$ at the bottom. Since the mean flow velocity at the sediment bed
is small, the volume averaged mean flow velocity is larger than that obtained without considering
the volume fraction term. Hence, \cite{schmeeckle14ns} underestimated the averaged fluid
velocity $\langle u \rangle$, and thus the friction coefficient $C_f$ calculated by using
Eq.~(\ref{eqn:frictionCoef}) is slightly larger.}

\begin{figure}[htbp]
  \centering
    \subfloat[]{
      \label{fig:sus-alpha}
      \includegraphics[width=0.45\textwidth]{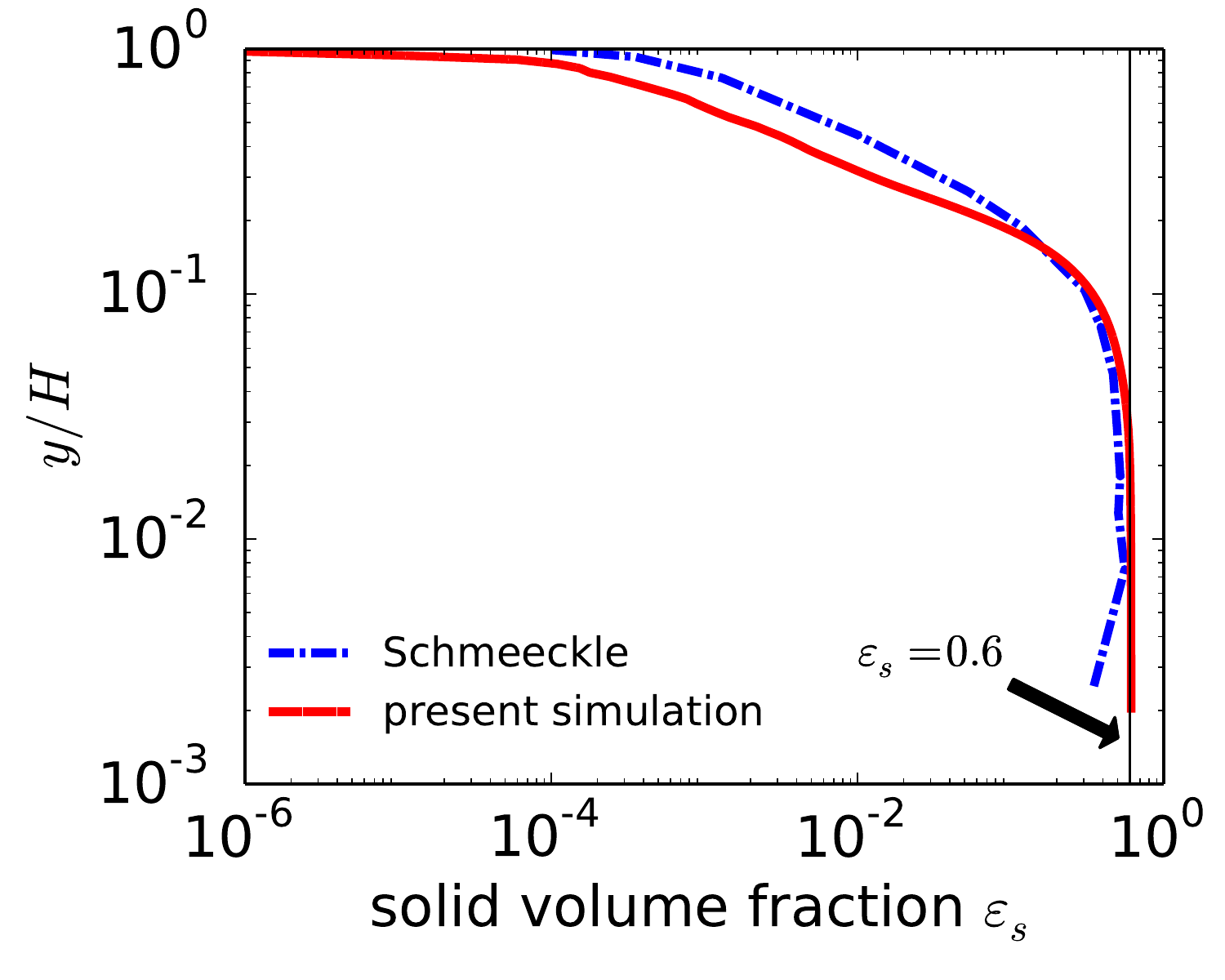}
    }
    \subfloat[]{
      \label{fig:sus-Ub}
      \includegraphics[width=0.45\textwidth]{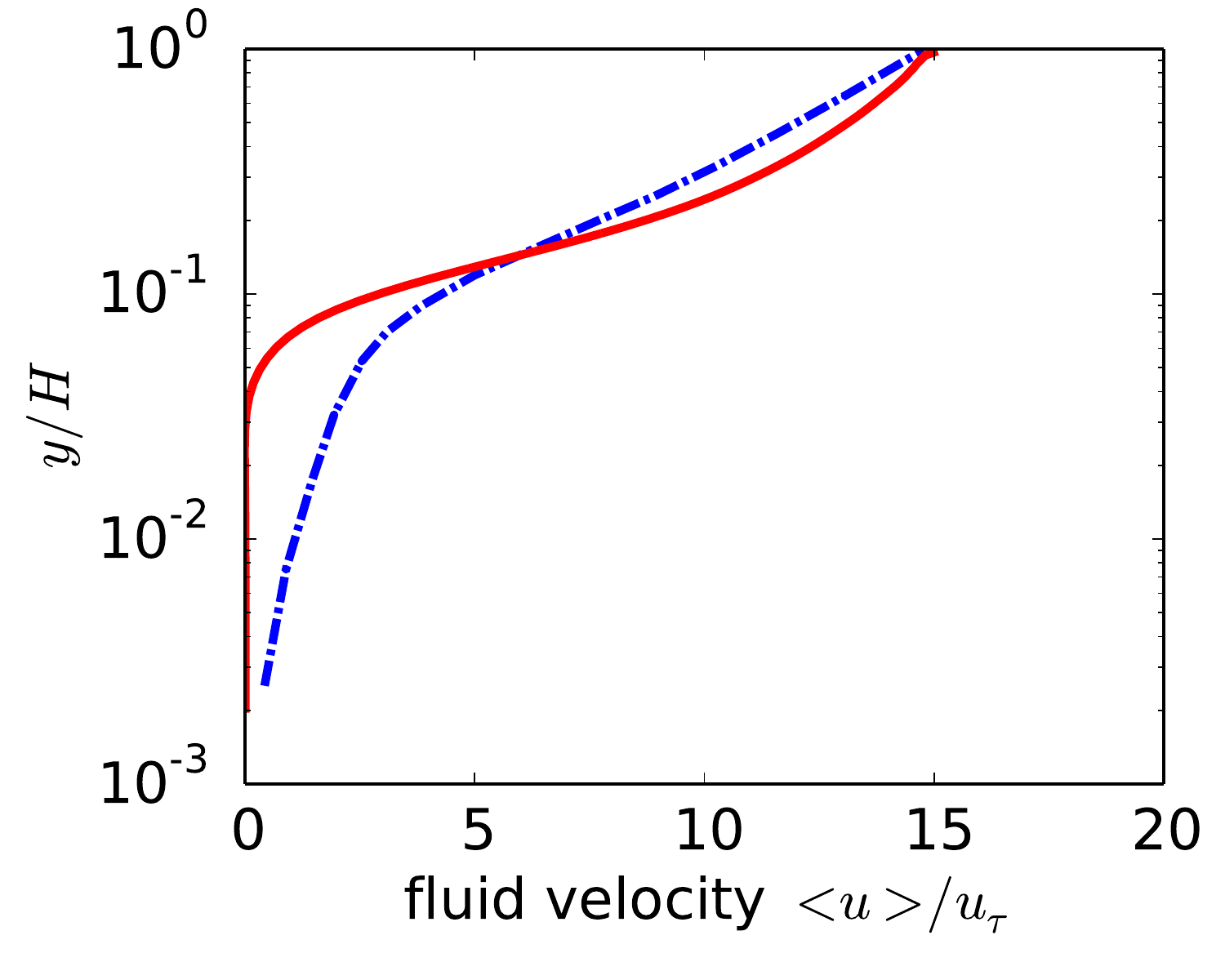}
    }
  \caption{The time-averaged properties of the suspend particles: (a) sediment volume fraction
  profile; (b) time-averaged velocity profile.}
  \label{fig:sus-alpha-vel}
\end{figure}

\begin{figure}[htbp]
  \centering
    \subfloat[]{
      \label{fig:sus-uu}
      \includegraphics[width=0.45\textwidth]{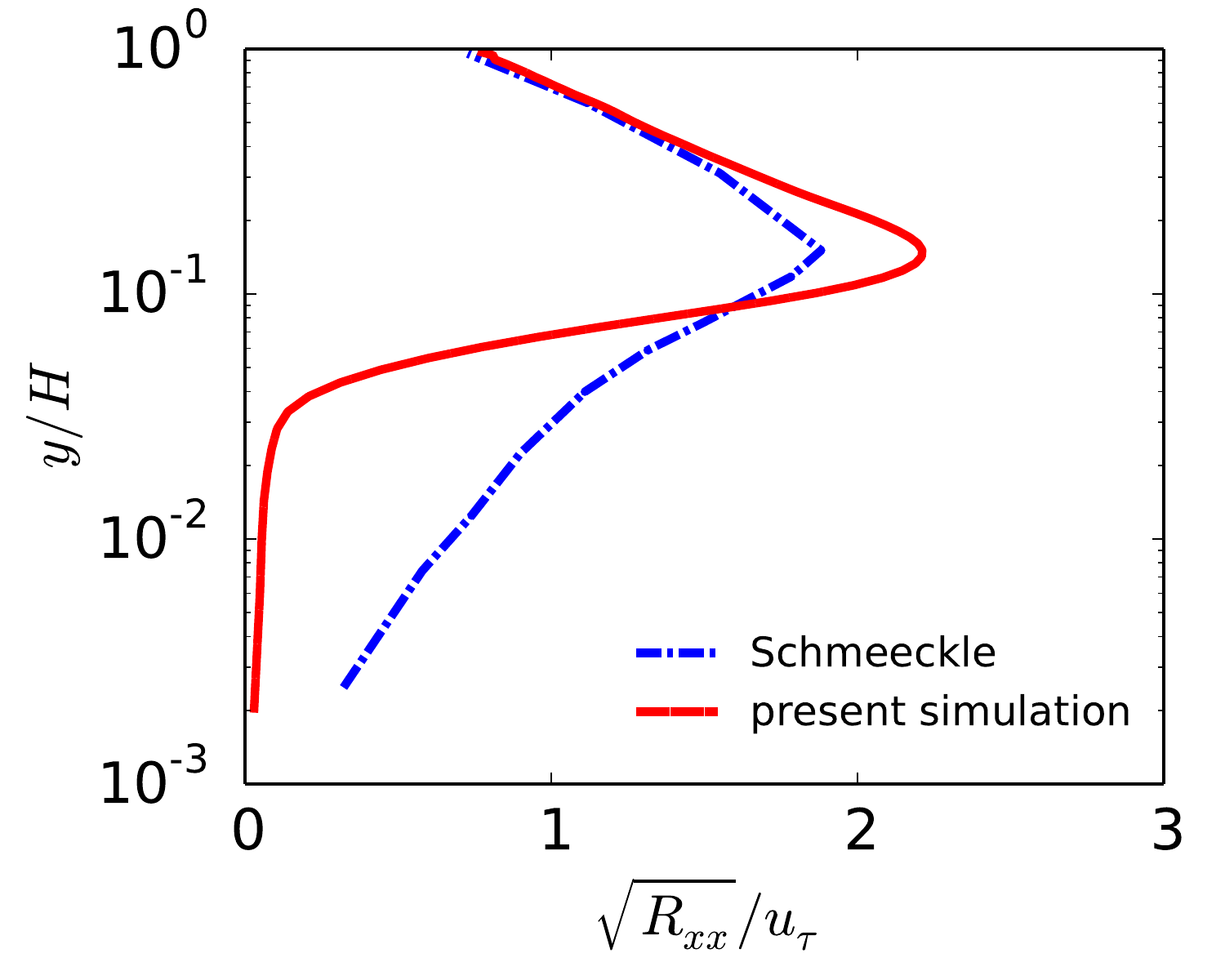}
    }
    \subfloat[]{
      \label{fig:sus-uv}
      \includegraphics[width=0.45\textwidth]{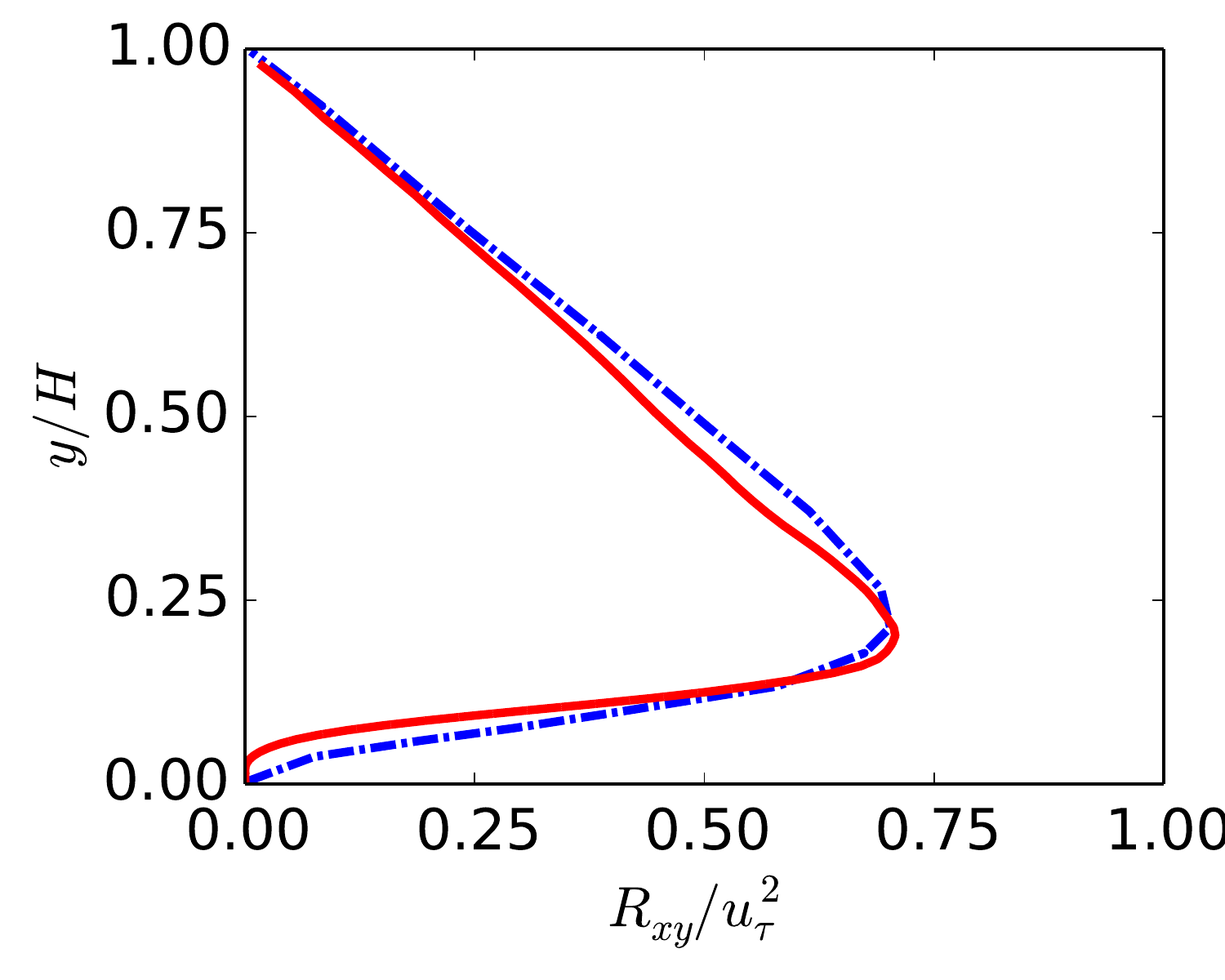}
    }
    \vspace{0.1in}
    \subfloat[]{
      \label{fig:sus-vv}
      \includegraphics[width=0.45\textwidth]{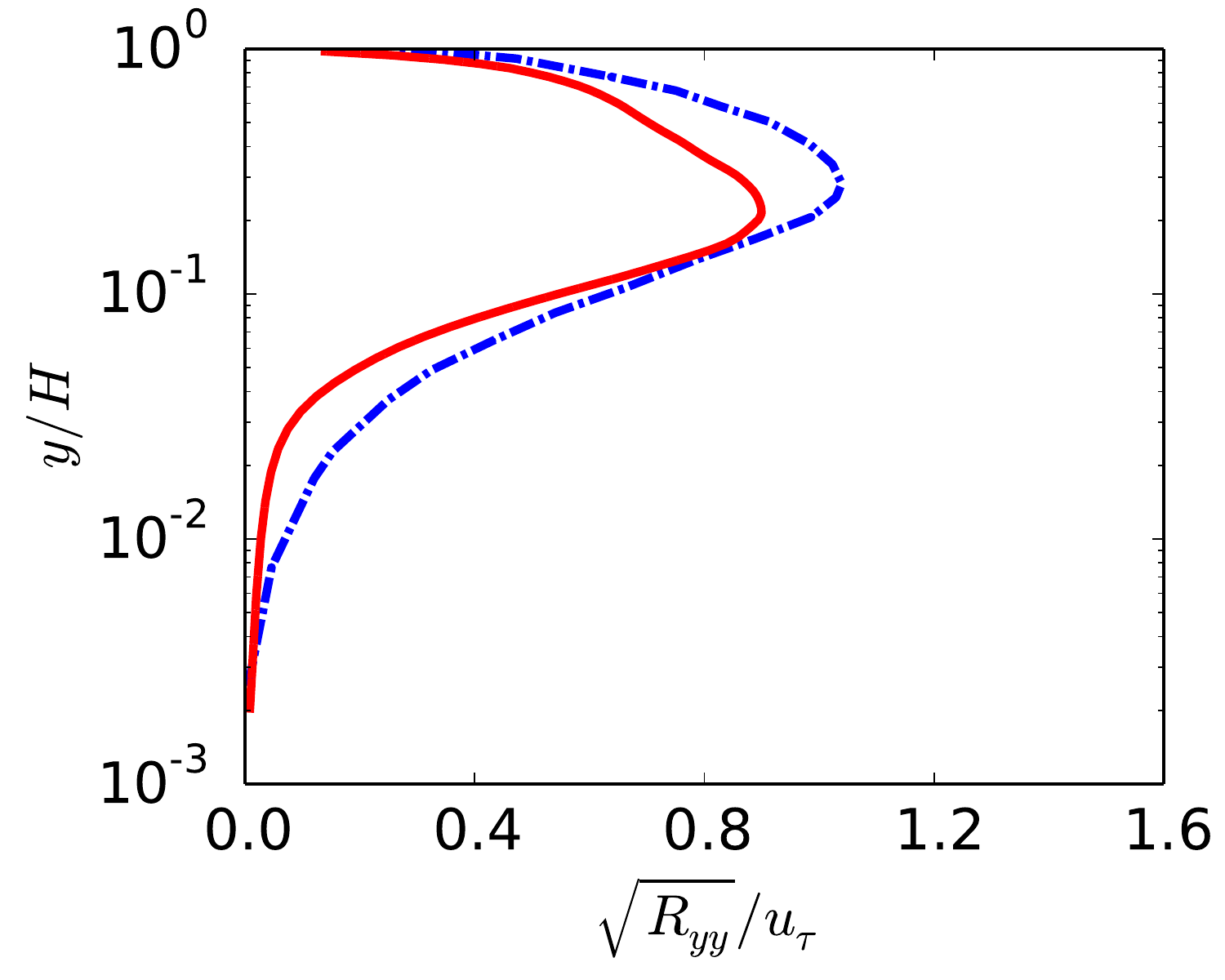}
    }
    \subfloat[]{
      \label{fig:sus-ww}
      \includegraphics[width=0.45\textwidth]{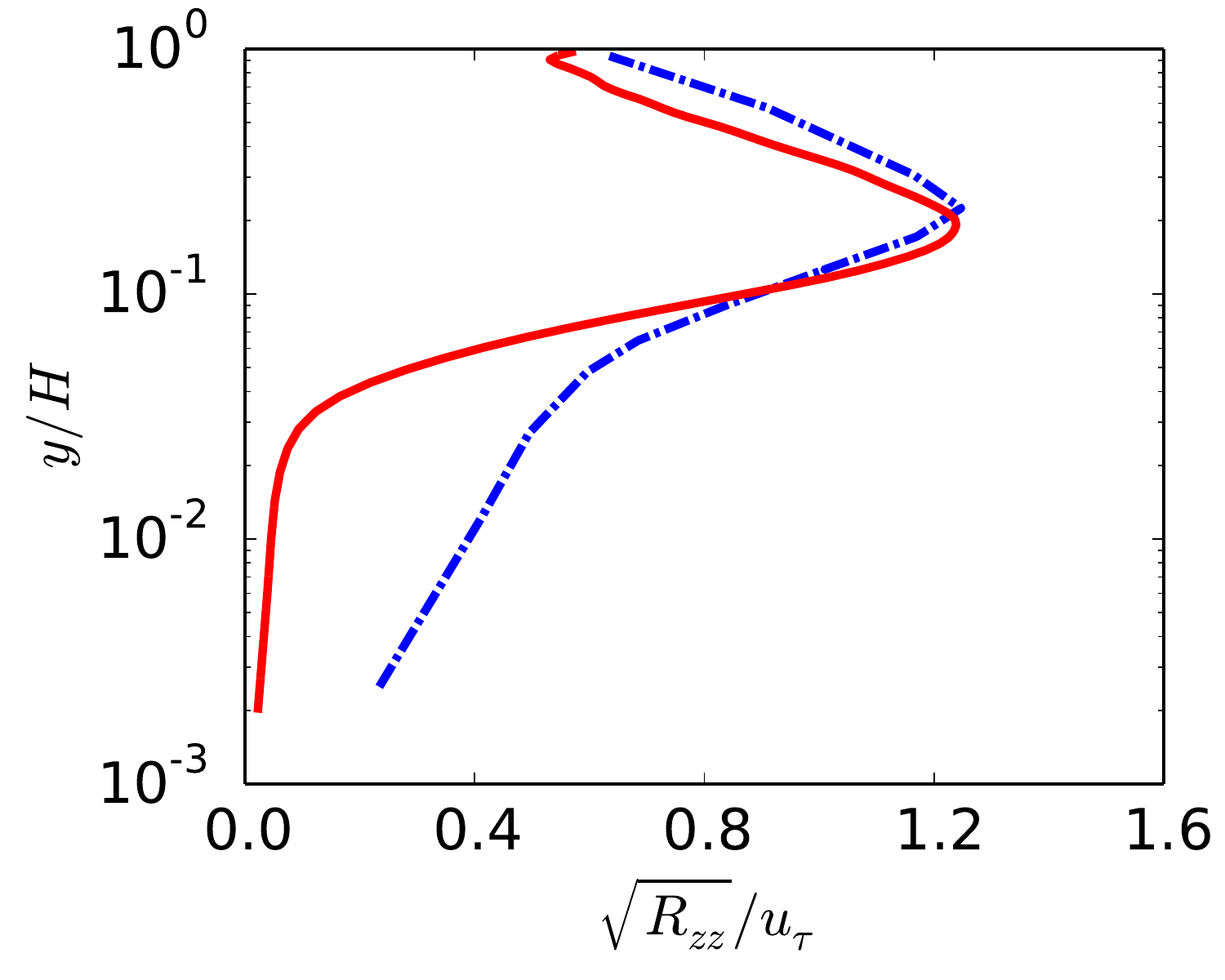}
    }
    \caption{Different components of the temporally and spatially averaged profiles of Reynolds
      stresses: (a) $R_{xx}$, (a) $R_{uv}$, (a) $R_{vv}$, (a) $R_{ww}$.}
  \label{fig:sus-Reynolds-stress}
\end{figure}

The temporally and spatially averaged profiles of sediment volume fraction and normalized fluid
velocity at $U_b = 1.2$~m/s are shown in Fig.~\ref{fig:sus-alpha-vel}(a)
and~\ref{fig:sus-alpha-vel}(b), respectively. It is noted that the solid volume fraction
($\varepsilon_s$) near the bottom obtained in the present simulation is about 0.6, which agrees
better with the experimental measurement~\citep{aussillous13io} than the results obtained
by~\cite{schmeeckle14ns}. In the present simulation, the diffusion-based averaging algorithm used
no-flux boundary condition to obtain the solid volume fraction $\varepsilon_s$ at the near-wall
region. When using no-flux boundary condition, mass conservation is guaranteed at the wall so that
the prediction of volume fraction $\varepsilon_s$ is more accurate. It can be seen from
Fig.~\ref{fig:sus-alpha-vel}(b) that the flow velocity obtained in the present simulation follows
the law of the wall as obtained in other cases by~\cite{schmeeckle14ns}. The immobile particle
boundary condition at the bottom provides more friction to the sediment particles so that the motion
of the bottom particles is constrained. Therefore, the velocity of the fluid flow is smaller due to
the drag force provided by the particles. The components of Reynolds stress are shown in
Fig.~\ref{fig:sus-Reynolds-stress}.  The discrepancies between the Reynolds stresses at the
near-wall region is because the bottom particles are fixed so that the flow fluctuation in the
present simulation is much smaller. The other turbulent shear stress components $\langle
u'w'\rangle$ and $\langle v'w'\rangle$ are very small and thus are not omitted in the figure.

\begin{figure}[htbp]
  \centering
  \includegraphics[width=0.8\textwidth]{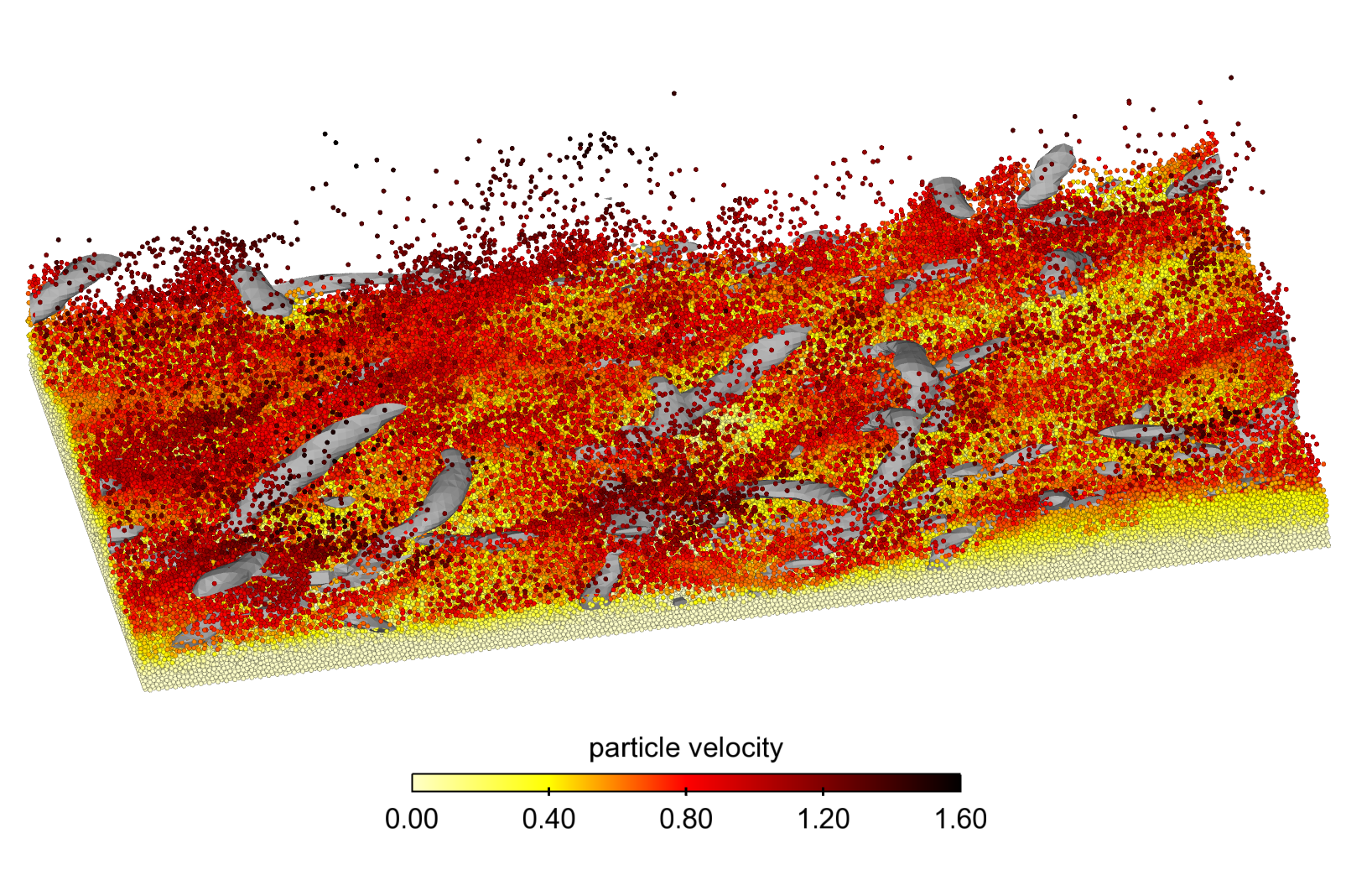}
  \caption{Vortical structures in the suspended particle transport using Q-criterion. The
  iso-surface of $Q = 20000$ is plotted, which is the second invariant of the velocity gradient
tensor. The unit of the particle velocity in the figure is m/s.}
  \label{fig:sus-Q}
\end{figure}

A snapshot of the iso-surface using Q-criterion is shown in Fig.~\ref{fig:sus-Q}, which
demonstrates the vortical structure in suspend sediment transport. It can be seen that the turbulent
eddies are observed at the fluid-particle interface, which is consistent with the results obtained
by~\cite{schmeeckle14ns}. Compared with the vortical structures in the Case 2b, the vortices in
suspend load regime are independent from the patterns of the sediment bed. This is because no sediment
dunes are generated to change the characteristics of the vortices.

\section{Discussions}
\label{sec:discussions}
{\color{black}
The proof-of-concept study in Section~\ref{sec:simulations} aims to demonstrate that CFD--DEM is
able to reproduce the integral or macroscopic quantities of ripple formation and morphological
evolution (e.g., wave length, dune height, evolution speed). This validation against experimental
data is a prerequisite for performing detailed physical interpretation of the simulation results.
Without that, we could risk being misled by numerical artifact of current simulations. However, the
investigation of the mechanics in sediment transport is the ultimate goal, and we took advantage of
the present CFD--DEM model to investigate the physical insights of sediment transport of different
regimes. The discussions on incipient motion in bedload, transition from bedload to suspended load,
and coexistence of bedload and suspended load are detailed below.

\subsection{Interpretations of Parameters Related to Particle Incipient Motion}
The critical Shields parameter is defined to describe the criteria for the incipient motion of
sediment particles. This value can be determined by employing visual observation or video imaging
techniques~\citep{smith04init} and varies at different Galileo
numbers~\citep{nielsen92cb,brownlie82pf}. However, the critical Shields stress is not easy to define
in terms of the sediment flux $q_i$. This is because sediment flux rate decreases gradually and will
not totally vanish when the Shields stress decreases, which is supported by both experimental
measurement and numerical simulation~\citep{smith04init,kidanemariam14dn}. The relationship of the
Shields stress and the sediment flux in Case 1 is shown in Fig.~\ref{fig:shields-1} as an example.
The critical Shields stress $\Phi_{Pois}$ in Poiseuille flow from experimental observation is
$0.12\pm0.03$~(Aussillous et al., 2013).  However, it can be seen in the figure that there is no
sudden change in the sediment flux near the critical Shields stress for both CFD--DEM and DNS
simulations. The results obtained in our simulations are consistent with previous
findings~\citep{smith04init,kidanemariam14dn} that it is difficult to define the minimum flux
$q_{min}$ for incipient sediment motion.

 \begin{figure}[!htpb]
 \centering
    \includegraphics[width=0.45\textwidth]{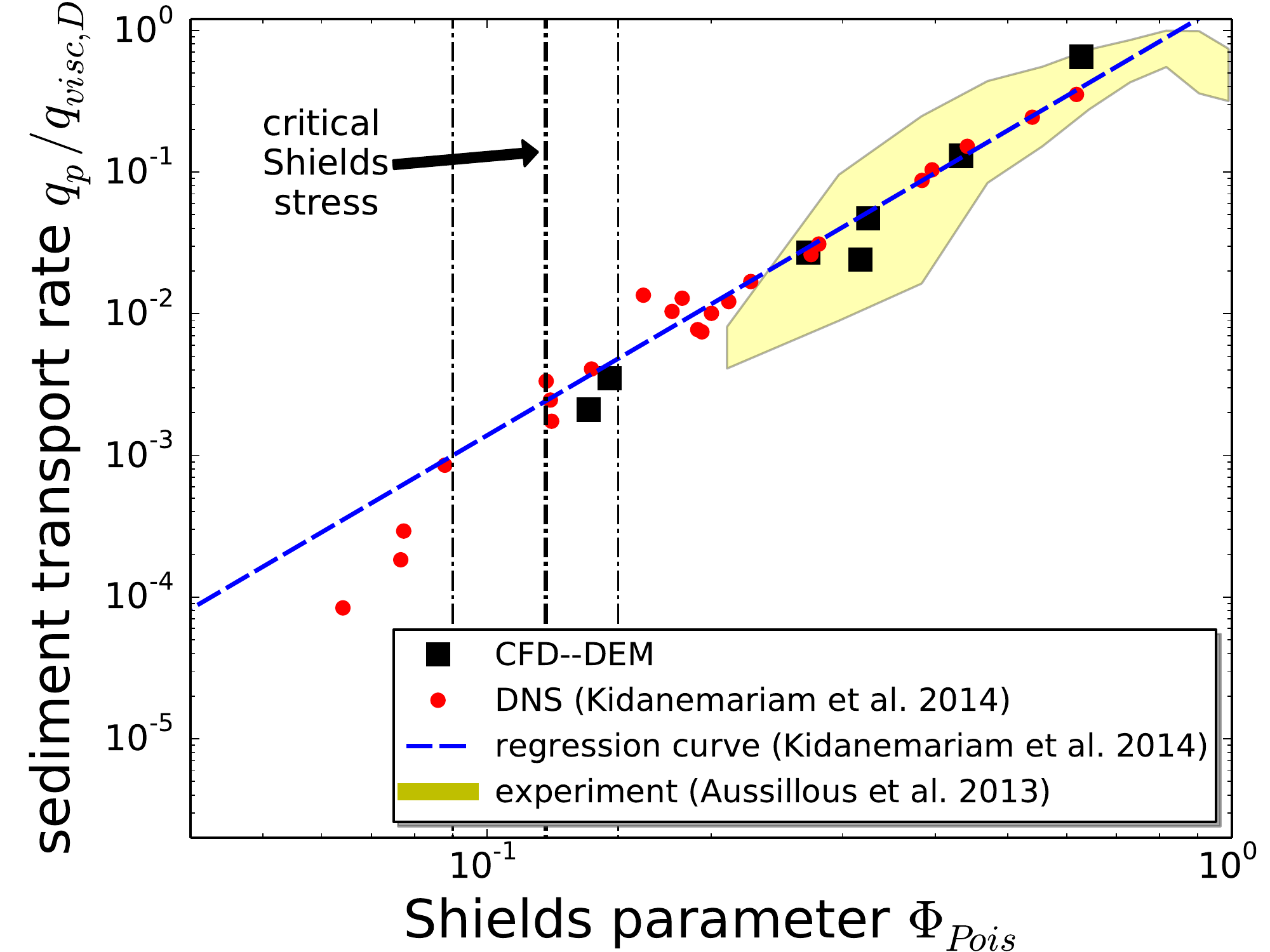}
 \caption{The sediment transport rate of Case 1 plotted as a function of the Shields parameter. The
 vertical dash-dot lines indicate the range of the Shields parameter.}
 \label{fig:shields-1}
 \end{figure}

\subsection{Transition from Bedload to Suspended Load}
The Bagnold criterion for suspension~\citep{bagnold66aa} denotes the threshold for the transition
from bedload to suspended sediment transport. To study this transition, numerical simulations are
performed based on the setup of Case 3 by using flow velocities ranging from 0.3~m/s to 1.2~m/s. The
sediment transport rate is plotted as a function of the friction velocity in
Fig.~\ref{fig:regime-bag}. It can be seen in the figure that the particles become suspended when the
friction velocity at the sediment bed is larger than fall velocity. This is consistent with Bagnold
criterion for suspension. Moreover, according to the observation from present simulations, the
transition from bedload transport to suspended load is not abrupt but gradual. In bedload regime,
when the friction velocity is approaching the Bagnold criterion, dunes are observed. If the friction
velocity further increases, the dunes first grow, and then gradually disappear due to the erosion of
flow.  When the friction velocity is larger than the fall velocity, particles become suspended.

\begin{figure}[!htpb]
\centering
\includegraphics[width=0.50\textwidth]{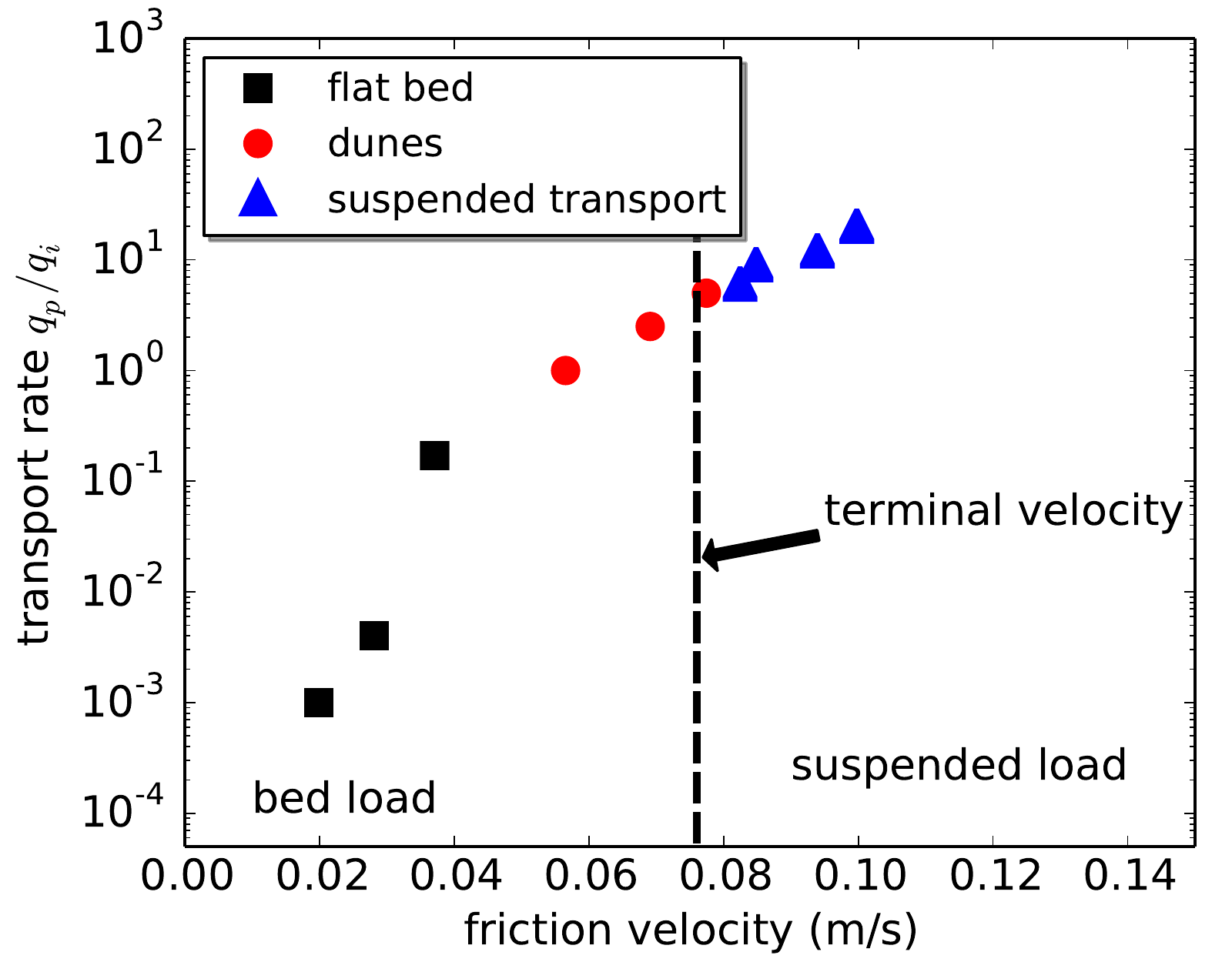}
\caption{The regime of sediment transport obtained by using CFD--DEM. The vertical line is drawn
according to Bagnold criterion.}
\label{fig:regime-bag}
\end{figure}

\subsection{Coexistence of Bedload and Suspended Load}
Calculations of sediment transport in engineering practice have assume either bedload or suspended
load depending on which mode is dominating. Whether or to what extent the two transport modes can
co-exist is an open question that is subject to debate. CFD--DEM simulations have the potential to
shed light on this issue. A typical snapshot taken from the Case 3 with $U_b = 1.2 m/s$ is presented
in Fig.~\ref{fig:bed-sus}(a), with the bedload and suspended load separated based on two different
criteria based on particle velocity (panel b) and particle concentration (panel c) of two threshold
values.  It can be seen in the figure that there is a layer of approximately $10d_p$ of particles
moving slowly as bedload at high Shields parameter. In our study, we use a threshold of particle
velocity $u_x < 3u_{term}$ to separate bedload from suspended load according to the maximum particle
velocity in bedload transport~\citep{schmeeckle14ns}, where particle terminal velocity $u_{term} =
0.077$~m/s. The threshold of the solid volume fraction $\varepsilon_s$ is also used to capture
bedload sediment transport, which is shown in Fig.~\ref{fig:bed-sus}(c). Indeed, the figure suggests
that the specific fractions of bedload and suspended load depend on the criterion used to delineate
them (i.e., based on particle velocity or particle volume fraction) and on the threshold values
(e.g., $epsilon_s = 0.1$~or~0.3). However, it is clear that regardless of the criterion or threshold
value adopted, bedload and suspended load co-exist in the snapshot analyzed, and both account for
significant portion of the total sediment flux.  

The empirical formulas calibrated on experiments primarily consisting of bedload can be very
inaccurate when used to predict flows with suspended load or a mixture of the transport modes, vice
versa for formulas developed for suspended load. This is illustrated in Fig.~\ref{fig:MPM}, which
shows revisions of the formula of \cite{meyer48fb} obtained by~\cite{wong06rc}
and~\cite{nielsen92cb} are applied to predict the sediment transport rate at different regimes. The
prediction of the sediment transport rate by the revised equation proposed by~\cite{wong06rc} is
based on the bedload and is significantly smaller than the prediction of~\cite{nielsen92cb}. To
investigate the differences between different revisions of the Meyer-Peter and M\"uller formula, we
separated bedload and suspended load using the threshold particle velocity $3 u_{term}$ and plotted
them as a function of the Shields parameter in Fig.~\ref{fig:MPM}. It can be seen from
Fig.~\ref{fig:MPM} that the bedload agrees with the formula proposed by~\cite{wong06rc}, and the
suspended load agrees with the formula proposed by~\cite{nielsen92cb}.  Therefore, the deviation of
the coefficient in different revisions of Meyer-Peter and M\"uller formula is due to the significant
increase of sediment transport rate from suspended load.

 \begin{figure}[!htpb]
 \centering
 \includegraphics[width=0.95\textwidth]{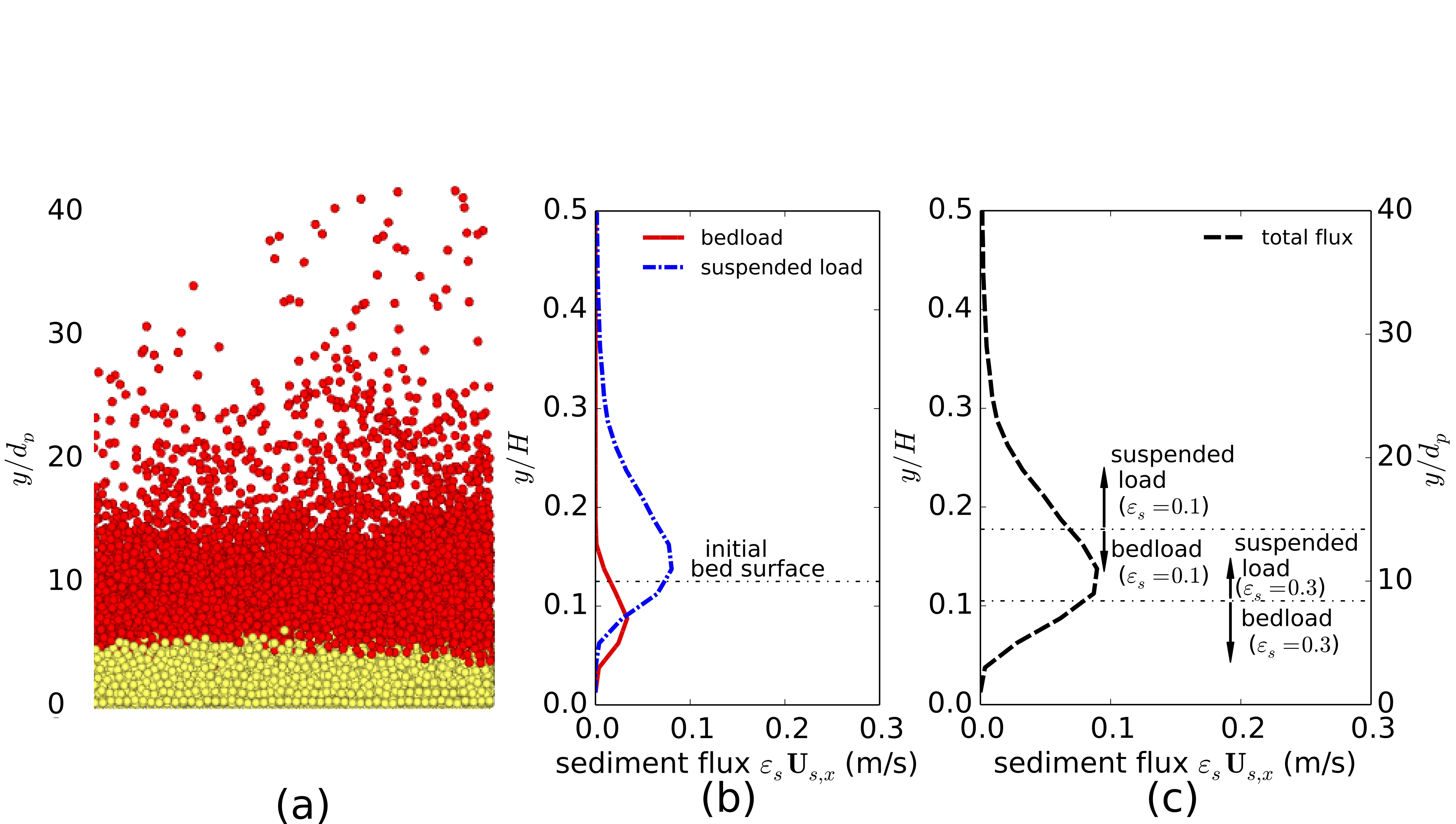}
 \caption{Comparison of bedload and suspended load in Case 3 at $U_b = 1.2$~m/s. Particles moving at
   $u_x > 3u_{term}$ are considered as suspended particle and colored by red in the left panel;
   particles moving at $u_x \le 3u_{term}$ are considered as bedload and colored by yellow in the
   left panel. Panel (a) and (b) are the comparisons of the vertical profiles of bedload flux and
   suspended load flux. The middle panel uses a threshold particle velocity at $u_x \le 3u_{term}$
   to capture the bedload; the right panel uses threshold solid volume fraction values of 0.1 and
   0.3. The initial sediment bed is approximately 10$d_p$, which correponds to $y/H = 0.125$.}
 \label{fig:bed-sus}
 \end{figure}

 \begin{figure}[!htpb]
 \centering
 \includegraphics[width=0.50\textwidth]{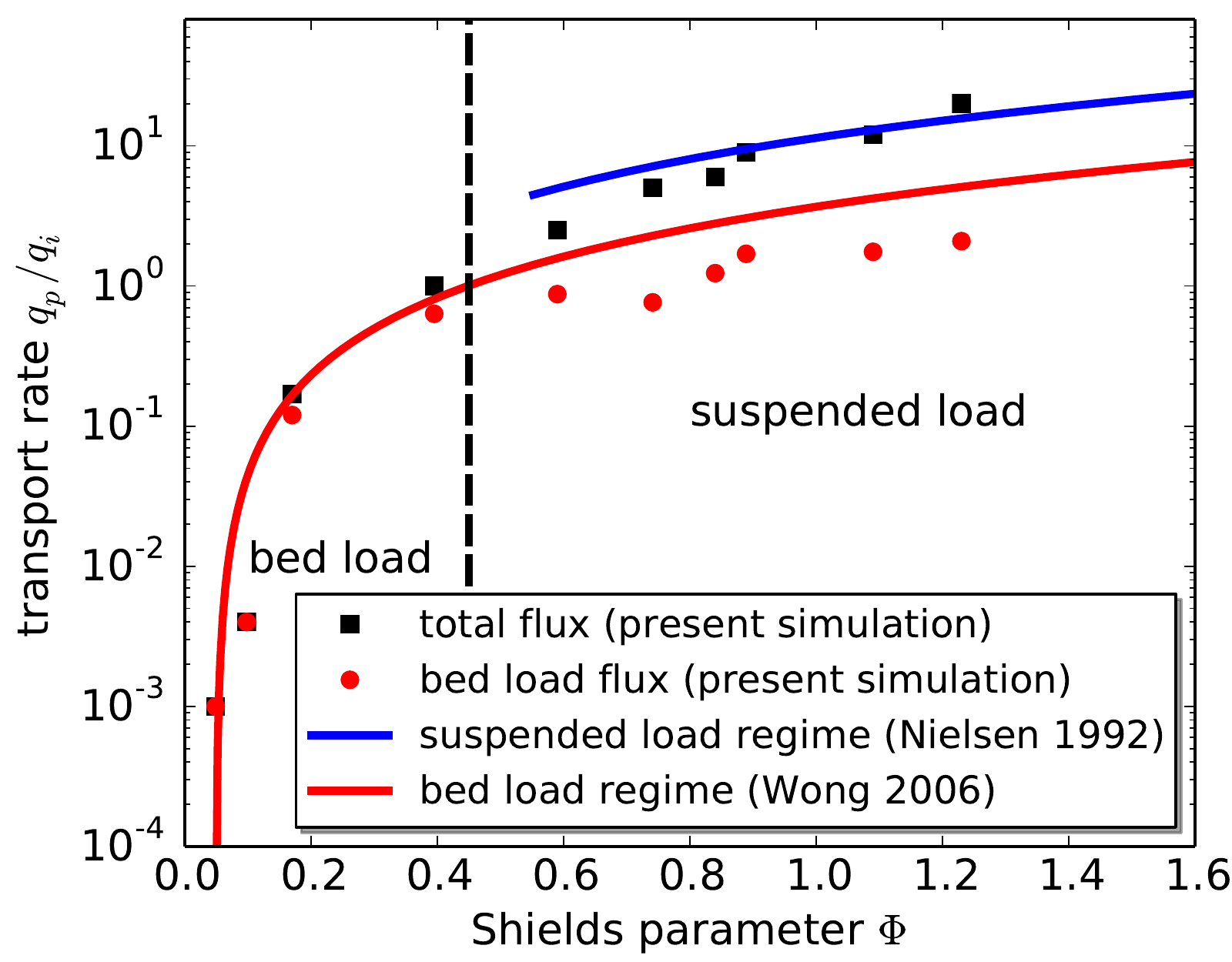}
 \caption{Comparison of the bedload flux and total flux in the present simulations. The particles
   moving at $u_x > 3u_{term}$ are considered as suspended particle, where $u_{term} = 0.077$~m/s is
   the terminal velocity of the sediment particle.}
 \label{fig:MPM}
 \end{figure}
 }

\section{Conclusion}
\label{sec:conclude}

In this work, a comprehensive study of current-induced sediment transport in a wide range of regimes
is performed by using CFD--DEM solver \textit{SediFoam}. Detailed quantitative comparisons are
performed using the results obtained in the present simulations and those in the literature. It is
demonstrated from the comparison that the accuracy of CFD--DEM is satisfactory for the simulation of
different sediment bed patterns. Considering the computational cost of CFD--DEM is much smaller than
that of the interface-resolved method, CFD--DEM is promising in the simulation of sediment
transport. This opens up the possibility to apply CFD--DEM to investigate realistic sediment
transport problems, e.g., the formation of ripples in the wave. 

{\color{black} In addition, the improvement of the results over existing CFD--DEM simulations is
demonstrated. We used a computational domain that is large enough to incorporate the bed form
(ripples), which is important advance over the featureless bed in the studies
by~\cite{schmeeckle14ns}. The second improvement of the present model is the averaging algorithm,
which enables mass conservation and resolves the boundary layer fluid flow simultaneously. Third, we
used a drag formulation that considered the influence of the volume fraction, which improves the
prediction of the sediment flux in suspended load. Finally, we considered the influence of
additional forcing terms in our numerical simulations, including added mass and lift force.} Because
of the improvements, the sediment transport rate in the suspended load regime agrees better with the
experimental results when the solid volume fraction is considered. Moreover, reasonable predictions
of the friction coefficient $C_f$ and the fluid flow at the sediment bed are reported.

\section{Acknowledgment}

The computational resources used for this project were provided by the Advanced Research Computing
(ARC) of Virginia Tech, which is gratefully acknowledged. We thank Dr. Kidanemariam for the
discussion that helped the numerical simulations in the present paper.  We thank the anonymous
reviewers for their comments, which helped improve the quality of the manuscript.  The authors
gratefully acknowledge partial funding of graduate research assistantship from the Institute for
Critical Technology and Applied Science (ICTAS, Grant number 175258).

\appendix
\end{document}